\documentclass[twocolumn]{aastex63}
\usepackage{comment}
\usepackage{amsmath}
\usepackage{rotating}
\usepackage{layouts}
\usepackage{tasks}
\usepackage{enumitem}
\usepackage{placeins}

\shorttitle{Spectrum of VHS 1256 b}
\shortauthors{Hoch et al.}

\begin{document}

\title{Moderate-Resolution $K$-Band Spectroscopy of the Substellar Companion VHS 1256 b}

\author[0000-0002-9803-8255]{Kielan K. W. Hoch}
\affiliation{Center for Astrophysics and Space Sciences,  University of California, San Diego, La Jolla, CA 92093, USA}

\author[0000-0002-9936-6285]{Quinn M. Konopacky}
\affiliation{Center for Astrophysics and Space Sciences,  University of California, San Diego, La Jolla, CA 92093, USA}

\author[0000-0002-7129-3002]{Travis S. Barman}
\affiliation{Lunar and Planetary Laboratory, University of Arizona, Tucson, AZ 85721, USA}

\author[0000-0002-9807-5435]{Christopher A. Theissen}
\altaffiliation{NASA Sagan Fellow}
\affiliation{Center for Astrophysics and Space Sciences, University of California, San Diego, La Jolla, CA 92093, USA}

\author[0000-0003-4297-7306]{Laci Brock}
\affiliation{Lunar and Planetary Laboratory, University of Arizona, Tucson, AZ 85721, USA}

\author[0000-0002-3191-8151]{Marshall D. Perrin}
\affiliation{Space Telescope Science Institute, 3700 San Martin Dr, Baltimore, MD 21218, USA}

\author[0000-0003-2233-4821]{Jean-Baptiste Ruffio}
\affiliation{Department of Astronomy, California Institute of Technology, Pasadena, CA 91125, USA}

\author[0000-0003-1212-7538]{Bruce Macintosh}
\affiliation{Kavli Institute for Particle Astrophysics and Cosmology, Stanford University, Stanford, CA 94305, USA}

\author[0000-0002-4164-4182]{Christian Marois}
\affiliation{NRC Herzberg Astronomy and Astrophysics, 5071 West Saanich Rd, Victoria, BC V9E 2E7, Canada}

\correspondingauthor{Kielan K. W. Hoch}
\email{kwilcomb@ucsd.edu}


\begin{abstract}
    We present moderate-resolution ($R\sim4000$) $K$ band spectra of the planetary-mass companion VHS 1256 b. The data were taken with the OSIRIS integral field spectrograph at the W.M. Keck Observatory. The spectra reveal resolved molecular lines from H$_{2}$O and CO. The spectra are compared to custom $PHOENIX$ atmosphere model grids appropriate for young, substellar objects. We fit the data using a Markov Chain Monte Carlo forward modeling method. Using a combination of our moderate-resolution spectrum and low-resolution, broadband data from the literature, we derive an effective temperature of 1240 K, with a range of 1200--1300 K, a surface gravity of $\log{g}=$ 3.25, with a range of 3.25--3.75 and a cloud parameter of $\log P_{cloud}=$ 6, with a range of 6.0--6.6. These values are consistent with previous studies, regardless of the new, larger system distance from GAIA EDR3  (22.2$^{+1.1}_{-1.2}$ pc). We derive a C/O ratio of 0.590$_{-0.354}^{+0.280}$ for VHS 1256b. Both our OSIRIS data and spectra from the literature are best modeled when using a larger 3 $\mu$m grain size for the clouds than used for hotter objects, consistent with other sources in the L/T transition region. VHS 1256 b offers an opportunity to look for systematics in the modeling process that may lead to the incorrect derivation of properties like C/O ratio in the high contrast regime.

\end{abstract}

\section{Introduction}

Direct imaging of exoplanets has revealed a population of Jupiter-like objects that orbit at wide separations \citep[$\sim$10--100 au;][]{bowler2016,nikolov2018,vigan2021}. 
In some cases, massive Jovians are found at separations of hundreds of au \citep[e.g.,][]{bailey2013}.
This population of widely-separated giants remains a puzzle for planet formation models. The orbits are too large for conventional core accretion models to make massive planets, while large-scale disk instabilities would create many more brown dwarf companions at wide separations than have been discovered in surveys around young stars \citep[e.g.,][]{rameau2013,biller2013,galicher2016,nielsen2019,vigan2021}.

Directly imaged exoplanets are difficult to observe in detail because they are faint compared to their bright host stars. Young, free floating and/or widely separated substellar and planetary mass objects often share the same color space as directly imaged gas giants \citep{faherty2013,liu2016}. These objects are extremely useful because of the ease in observation and insight they provide into atmospheric properties of gas giant exoplanets. 

 Formation of molecules and condensates are the dominant opacity sources of the atmospheres of giant planets and brown dwarfs. Prominent molecules include H$_2$O, H$_2$, CO, CH$_4$, NH$_3$, CrH, FeH, and CaH and their absorption features establish much of the L and T spectral shapes. As these materials condense in the atmospheres, clouds are expected to form. Clouds are a nearly ubiquitous feature of substellar objects, but have remained difficult to model. The presence of clouds is a function of both temperature and surface gravity, which makes directly imaged companions, that can span the L/T transition, excellent candidates for cloud characterization.

Brown dwarfs have similar spectral features as young directly imaged planets, and therefore are excellent test cases for verifying modeling procedures that result in atmospheric characterization. In particular, brown dwarfs that are members of binary systems are extremely useful, as they offer a means to verify properties such as metallicity if their companion is stellar \citep{wang2022}. Additionally, since they formed from the same material, brown dwarfs that form in binary systems of two substellar companions should at least have the same relative composition. Many brown dwarfs also have existing constraints on effective temperature, surface gravity, and composition. Processes that impact the atmospheric abundances such as mixing and condensation vary with effective temperature and gravity, so it is important for those parameters to be constrained with independent data.

VHS J125601.92-125723.9b (VHS 1256 b, \citealt{gauza2015}) is a wide orbit (8'' separation) substellar companion to the late M dwarf binary star system VHS 1256-1257 AB \citep{stone2016}.  The age of this system has not been determined, but weak absorption from neutral gases (Na and K) and collision-induced absorption by H$_2$ detected in VHS 1256 b's atmosphere indicate low surface gravity and therefore suggests a young age \citep{gauza2015}. Gaia EDR3 measured a new parallax for VHS 1256 AB b which yielded a new distance, 22.2$^{+1.1}_{-1.2}$ pc, adjusted from the previous estimate 12.7$\pm$1.0 pc \citep{gauza2015}. This change in distance has the potential to affect the temperature and mass measurements of VHS 1256 b. This object is also highly variable and can cause some uncertainty when interpreting atmospheric parameters. The rotation ($v\sin i$) of VHS 1256 b was measured by \cite{bryan2018} to be $13.5^{+3.6}_{-4.1}$ km s$^{-1}$. The apparent redness of this object and previous modeling of the optical to mid infrared photometric data suggests that VHS 1256 b has a very cloudy atmosphere \citep{rich2016}. \cite{gauza2015} estimated that VHS 1256 b is an L7.0$\pm$1.5 object, right in the L/T transition. \cite{miles2018} presented $L$-band spectra of the object that shows signs of weak methane absorption and a thick cloudy atmosphere. This companion is also a target in an ERS program for the James Webb Space Telescope \citep{hinkley2017}.  A summary of the VHS 1256 system parameters and photometry is given in Table \ref{tab:sum_system}.

Here, we present $R\sim4000$ $K$-band spectra of VHS 1256 b using the OSIRIS instrument on the W.M. Keck Telescope. In Section 2 we report our observations and data reduction methods. In Section 3 we use atmosphere model grids and forward modeling Markov Chain Monte Carlo methods to determine the best-fit effective temperature, surface gravity, and metallicity of both the VHS 1257b and VHS 1256AB. We use our best-fit parameters and $PHOENIX$ models with scaled molecular mole fractions to derive a C/O ratio for VHS 1256 b and basic atmospheric parameters for the host star. In Section \ref{sec:discussion} we discuss the implications of our results and future work. \par

\section{Data Reduction}\label{sec:data}
VHS 1256 b was observed on 2017 June 10 with the OSIRIS integral field spectrograph (IFS) on the W.M. Keck I telescope \citep{larkin2006}. We used the $K$ broadband mode (1.965--2.381 $\mu$m) with a spatial sampling of 20 milliarcseconds per lenslet. Since the primary star in the system is faint (V$\sim$18), we used the Keck I facility laser guide star adaptive optics system to achieve close to diffraction-limited performance \citep{chin2012}. The primary star was used as the tip/tilt reference (R$\sim$15), while the laser was used for all other corrections. We acquired the primary, and then offset to the location of the companion, since the OSIRIS field of view in 20 milliarcsecond mode is only $0.32\arcsec \times 0.28\arcsec$. In total, we integrated on the object for 70 minutes (7 exposures of 10 minutes each, where we dithered up and down by a between exposures). Observations of an A0V telluric standard (HIP 67139) were obtained close in time to the data. We also obtained dark frames with exposure times matching our dataset. We did not obtain separate sky exposures, electing instead to nod the object up and down on the detector for efficiency and then use the nod pairs for sky subtraction. The data were reduced using the OSIRIS data reduction pipeline \citep[DRP;][]{krabbe2004,lockhart2019}. Data cubes are generated using the standard method in the OSIRIS DRP, using rectification matrices provided by the observatory. At the advice of the DRP working group, we did not use the Clean Cosmic Rays DRP module (T. Do, priv. comm). We did not use scaled sky subtraction. 

\begin{figure*}
\epsscale{0.25}
\plotone{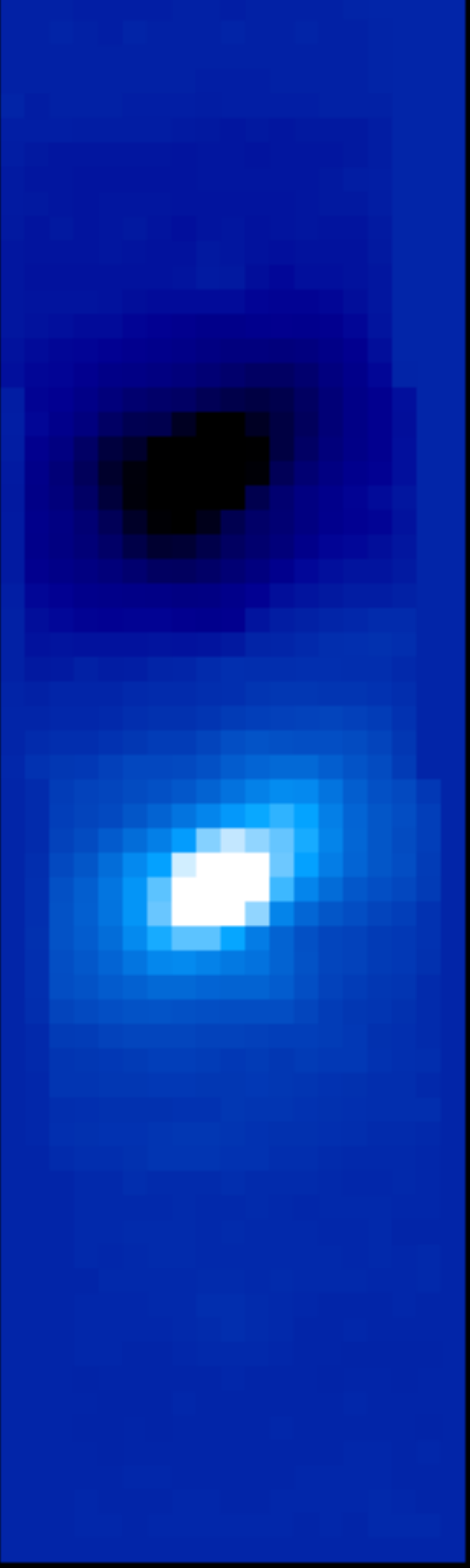}
\caption{An example data cube image frame, collapsed in wavelength via median, from our OSIRIS VHS 1256 b data set. The field of view is $0.32\arcsec$ x $1.28\arcsec$. The dark spot represents the negative value from the companion subtraction at the other nod position. The bright spot shows the companion clearly visible without any need for speckle removal.  The elongation of the PSF is due to the imperfect LGS AO correction, but does not impact the resulting extracted spectra, which are created by summing the spectra from multiple spaxels around the core of the PSF.}
\label{fig:osiris}
\end{figure*}

After extracting one-dimensional spectra for the telluric sources, we used the DRP to remove hydrogen lines, divide by a blackbody spectrum, and combine all standard star spectra for each respective night. Once the standard star spectra were combined, a telluric calibrator spectrum was obtained. The telluric correction for VHS 1256 b was then done by dividing the final combined telluric calibrator spectrum from all object frames.

Once the object data cubes are nod subtracted and fully reduced, we identify the location of the target. In high contrast observations with OSIRIS, the location of a companion can be challenging to find due to the brightness of the speckles. For VHS 1256 b, the starlight from the host star is not bright enough to impact the spectrum of the companion, and identification is straightforward (see Figure \ref{fig:osiris}). 

We extract the object spectrum using a box of $3 \times 3$ spatial pixels (spaxels). Once we extracted the VHS 1256 b spectra from each frame for all data, we then normalize each individual spectrum to account for PSF and background fluctuations. Finally, we median-combine all 7 individual spectra. To calibrate the flux of our spectra we calculated the flux at each wavelength such that, when integrated, the flux matches the most recent $K$-band apparent magnitude ($14.665 \pm 0.01$) from \cite{gauza2015}.

We note that the telluric correction in the blue portion of the spectrum appears to leave some residual telluric signal. In our previous work, simultaneous telluric correction was possible using the speckles from the early-type host stars \citep{wilcomb2020}. This is not the case for the well-isolated VHS 1256b. We attempt to mitigate the impact of the imperfect telluric removal in our subsequent analysis.

Uncertainties were determined by calculating the RMS between the individual spectra at each wavelength. These uncertainties include contributions from statistical error in the flux of the planet as well as some additional error in the blue end of the spectrum due to the residual telluric features in this region. The OH sky lines are well-subtracted and have a negligible contribution to the uncertainties.  The median uncertainty for our spectrum with the continuum included is $\sim$7$\%$.

 \begin{figure*}
\epsscale{0.95}
\plotone{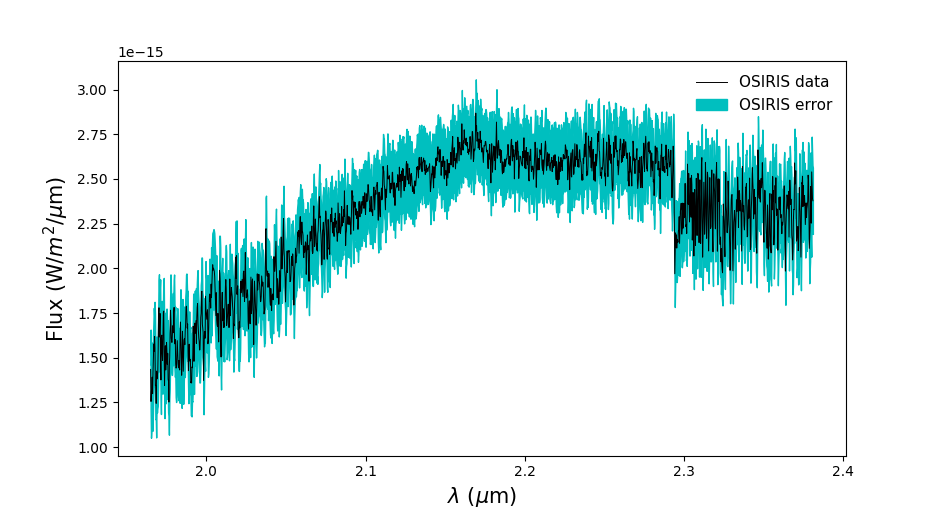}
\caption{Our fully reduced, combined, and flux calibrated moderate-resolution OSIRIS $K$-band spectra of VHS 1256 b. The errors are derived from the RMS of the individual spectra at each wavelength.  Such as error includes uncertainties in both the continuum and the lines, the former of which tends to be larger and is removed via continuum subtraction in portions of our analysis.  These total spectral uncertainties are represented as a shaded blue region.  The median uncertainty is $\sim$7$\%$ in flux. The CO bandhead is clearly visible at 2.3 $\mu$m.}
\label{fig:extracted_spec}
\end{figure*}
 
 By removing the continuum, narrow spectral features are more easily analyzed by avoiding low spatial frequency errors that impact the overall continuum shape. We remove the continuum from our fully reduced and flux calibrated spectra using a similar continuum removal strategy we have employed in the past for OSIRIS data (e.g., \citealt{barman2015,wilcomb2020}). To remove the continuum, we apply a high-pass filter with a kernel size of 200 spectral bins to each of the individual spectra. Then we subtract the smoothed spectrum from the original spectra. Once all the individual spectra are continuum subtracted, we median combine them and find the uncertainties by calculating the RMS of the individual spectra at each wavelength.  The median uncertainty on our flattened spectrum is 0.4$\%$, much smaller than the uncertainty on the spectrum with the continuum included.

\section{Spectral Modeling}
\subsection{Synthetic Spectra}

In order to determine the atmospheric properties of VHS 1256b, we use models computed through the $PHOENIX$ model atmosphere code \citep{hauschildt1999}. The atmospheric properties for young gas giants and brown dwarfs are notoriously complex, with cloud modeling being particularly challenging. In our modeling framework, we are able to parameterize clouds in several different ways. They are incorporated as a thick cloud layer with a log-normal ISM-like grain size distribution in all three types of cloud models; DUSTY \citep{allard2001}, COND \citep{allard2001}, and an Intermediate Cloud Model (ICM) \citep{barman2011}. DUSTY models have inefficient gravitational settling so dust is still in the atmosphere, and COND models have efficient gravitational settling therefore no dust is left in the atmosphere. For VHS 1256 b, we used an ICM approach that has a base pressure set by equilibrium chemistry. The vertical extent of the cloud is parameterized by a pressure value that, when above the value, allows the cloud particles number density to fall off exponentially. The cloud particle sizes follow a log normal distribution that is characterized by a median grain size. We will therefore refer to the custom grid constructed here as \textit{PHOENIX-ICM} to distinguish it from other models based on the \textit{PHOENIX} framework. We also used two other grids to fit the data; \textit{PHOENIX-ACES}, the custom grid used for $\kappa$ And b \citep{wilcomb2020}, and \textit{PHOENIX-HR8799}, the custom grid used in \cite{ruffio2021}. We explore the results of our choice of cloud model and describe the fits from a few other models in Section \ref{sec:discussion}.    

Our first goal is to constrain the temperature, surface gravity, and cloud parameters of VHS 1256 b. In order to do this, we must construct a model grid that spans the expected values of these parameters. In a number of previous works on VHS 1256 b, the temperature was estimated to be in the range of 800--1200 K and the surface gravity ($\log g$) 3.2--4.5 \citep[e.g.,][]{miles2018,gauza2015}.

Based on these measurements, we generated a custom \textit{PHOENIX} grid. The details on the computation of this grid are described in \cite{barman2011,barman2015}, with the updated methane line list from \cite{hargreaves2020} and the optical opacities from \cite{karkoschka2010}. The grid spans a temperature range 800--2100~K, a $\log g$ range of 3.5--5.5~dex, and a cloud property range $P_{cloud}$ dyne/cm$^2$ of 5$\times$10$^5$--4$\times$10$^6$, which encompasses the values previously reported for VHS 1256 b. Vertical mixing (Kzz) was modeled in \cite{miles2018} and was determined to be $\sim$ 10$^{8}$ cm$^2$s$^{-1}$ because with a cooler object of about 1200 K some of the C will still be in CH$_4$, and therefore CO will not be constant with height. This custom grid also varied grain size, but initially we used a fixed mean grain size of 0.5 $\mu$m.

The synthetic spectra from the grid were calculated with a wavelength sampling of 0.05~\AA\ from 1.4 to 2.4~$\mu$m. Each spectrum was convolved with a Gaussian kernel with a FWHM that matched the OSIRIS spectral resolution \citep{barman2011}. Both flux calibrated and continuum subtracted data were modeled and analyzed. The synthetic spectra were flux calibrated and continuum subtracted using the same routines as the data. 

\subsection{Forward Modeling}\label{sec:for_mod}

To determine the best-fit model for each grid, we use a forward-modeling approach following \cite{blake2010}, \cite{burgasser2016}, \cite{hsu2021}, and \cite{theissen2022} The effective temperature ($T_\mathrm{eff}$), surface gravity ($\log g$), and cloud parameter ($\log P_{cloud}$) are inferred using a Markov Chain Monte Carlo (MCMC) method built on the \texttt{emcee} package that uses an implementation of the affine-invariant ensemble sampler \citep{goodman2010,foreman-mackey2013}. The assumptions and description of the MCMC calculations are described in \cite{wilcomb2020}.

We also forward modeled the primary, VHS 1256 AB, spectra from \cite{gauza2015}. The data span 0.47--1.35 $\mu$m, 1.48--1.78 $\mu$m, and 1.95--2.41 $\mu$m.  Since they are a late-M binary pair, we used the G\"ottingen spectral library \citep{husser2013} that varies the effective temperature ($T_\mathrm{eff}$), surface gravity ($\log g$), and metallicity ([M/H]) and covers the range of values expected for an M star. The results show that the host binary has approximately solar metallicity. The best fit parameters for VHS1256AB from $K$-band data are $T_\mathrm{eff}$ = 2756$^{+13}_{-15}$ K, $\log g$ = 4.99$^{+0.0076}_{-0.013}$, and [M/H] = 0.0093$^{+0.012}_{-0.0063}$. For the short wavelength data the best fit parameters are $T_\mathrm{eff}$ = 2668$^{+8}_{-7}$ K, $\log g$ = 3.57$\pm0.060$, and [M/H] = 0.20$\pm0.040$. The $H$-band data best fit parameters are $T_\mathrm{eff}$ = 2758$^{+15}_{-18}$ K, $\log g$ = 3.92$^{+0.090}_{-0.40}$, and [M/H] = 0.39$^{+0.14}_{-0.35}$. Since there is no higher resolution data available for the host binary, our precision in metallicity or abundance measurements is limited and further discussion about the systematics in the uncertainties is in Section \ref{sec:temp_g_c}. We report the allowed values of these fits to be an effective temperature of 2661 -- 2773 K, a surface gravity of 3.51 -- 5.00, and an [M/H] of 0.0030 -- 0.53. The full results of the fits are shown in Table \ref{tab:primary}.

\subsection{Temperature, Gravity, and Cloud Parameters}\label{sec:temp_g_c}

\begin{figure*}
\epsscale{0.95}
\plotone{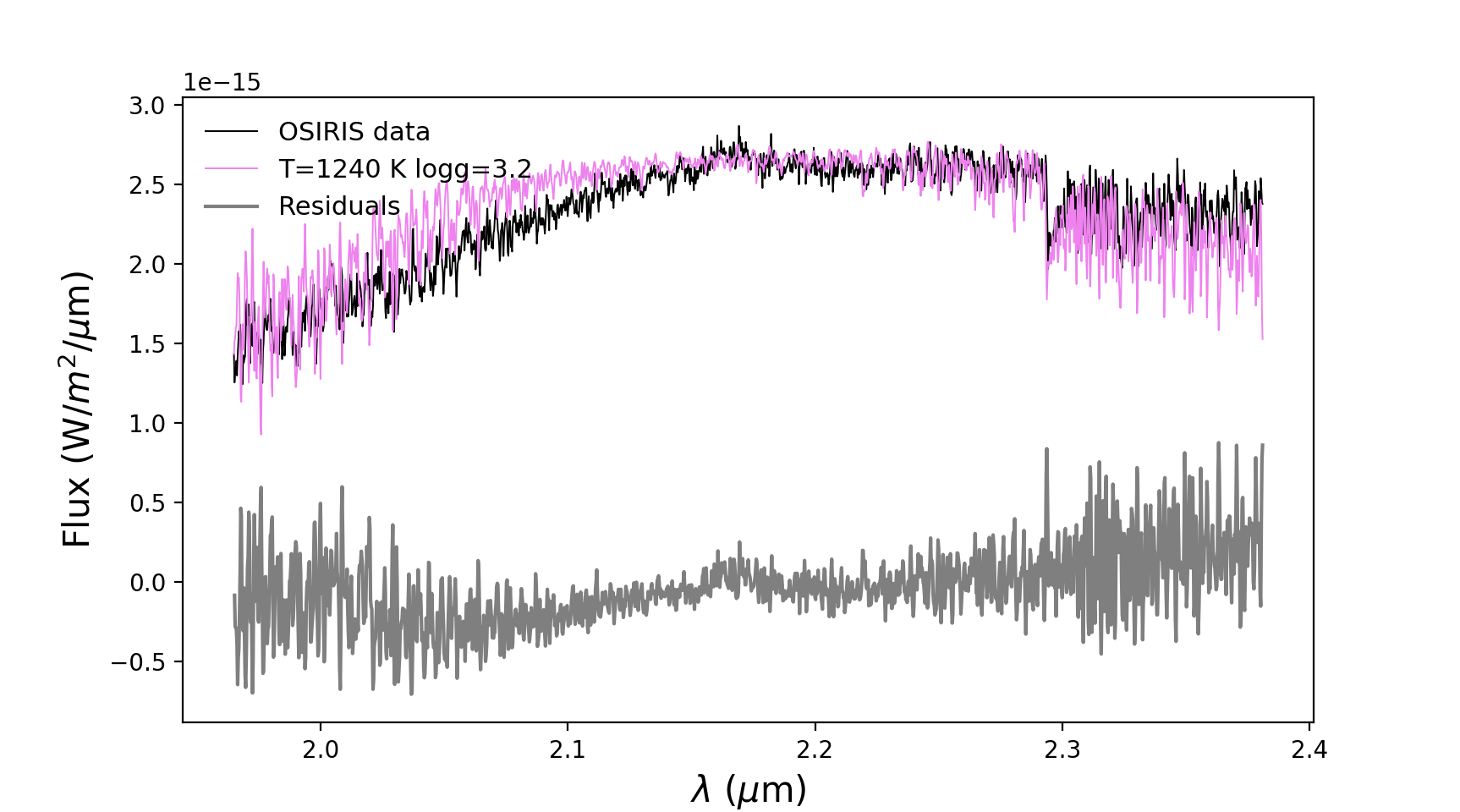}
\caption{Our fully reduced, combined, and flux calibrated moderate-resolution OSIRIS $K$-band spectra of VHS 1256 b in black plotted alongside our best-fit model in pink. The residuals are shown below in gray. Even our custom \textit{PHOENIX-ICM} had trouble fitting the continuum of our spectrum, specifically on the bluer end. However, this grid still provided the lowest residuals of any we considered.}
\label{fig:continuum_mod}
\end{figure*}

\begin{figure*}
\epsscale{0.95}
\plotone{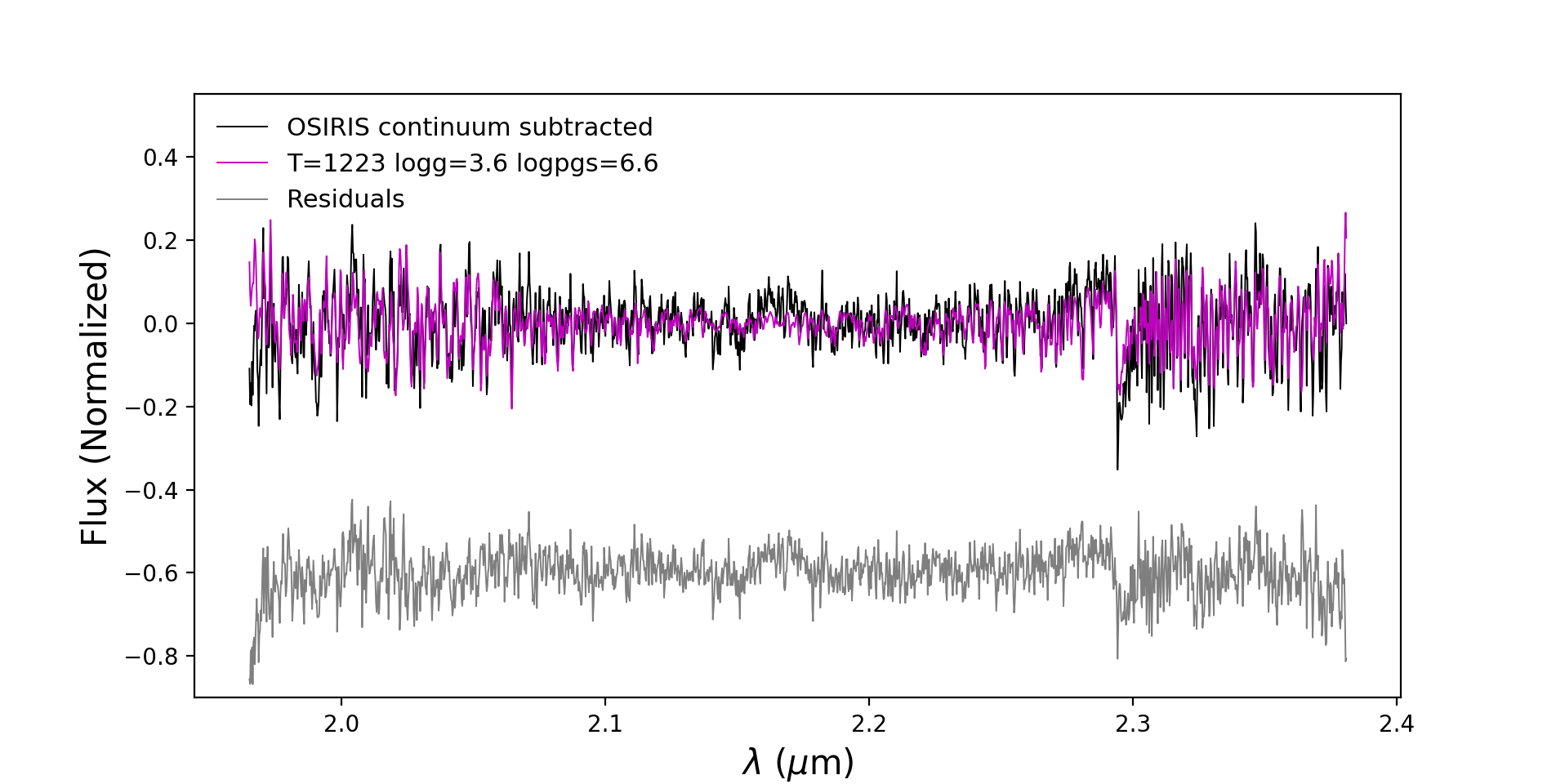}
\caption{Our OSIRIS continuum subtracted $K$-band spectra in black plotted against our MCMC best-fit model in magenta. The bump in the center of the spectrum between 2.1 and 2.2 $\mu$m is due to residual telluric. By removing the continuum, we remove uncertainties that were tied to the continuum, thus the residuals for the flatted fit are much smaller.}
\label{fig:flat_fit}
\end{figure*}

\begin{figure*}
\epsscale{0.95}
\plotone{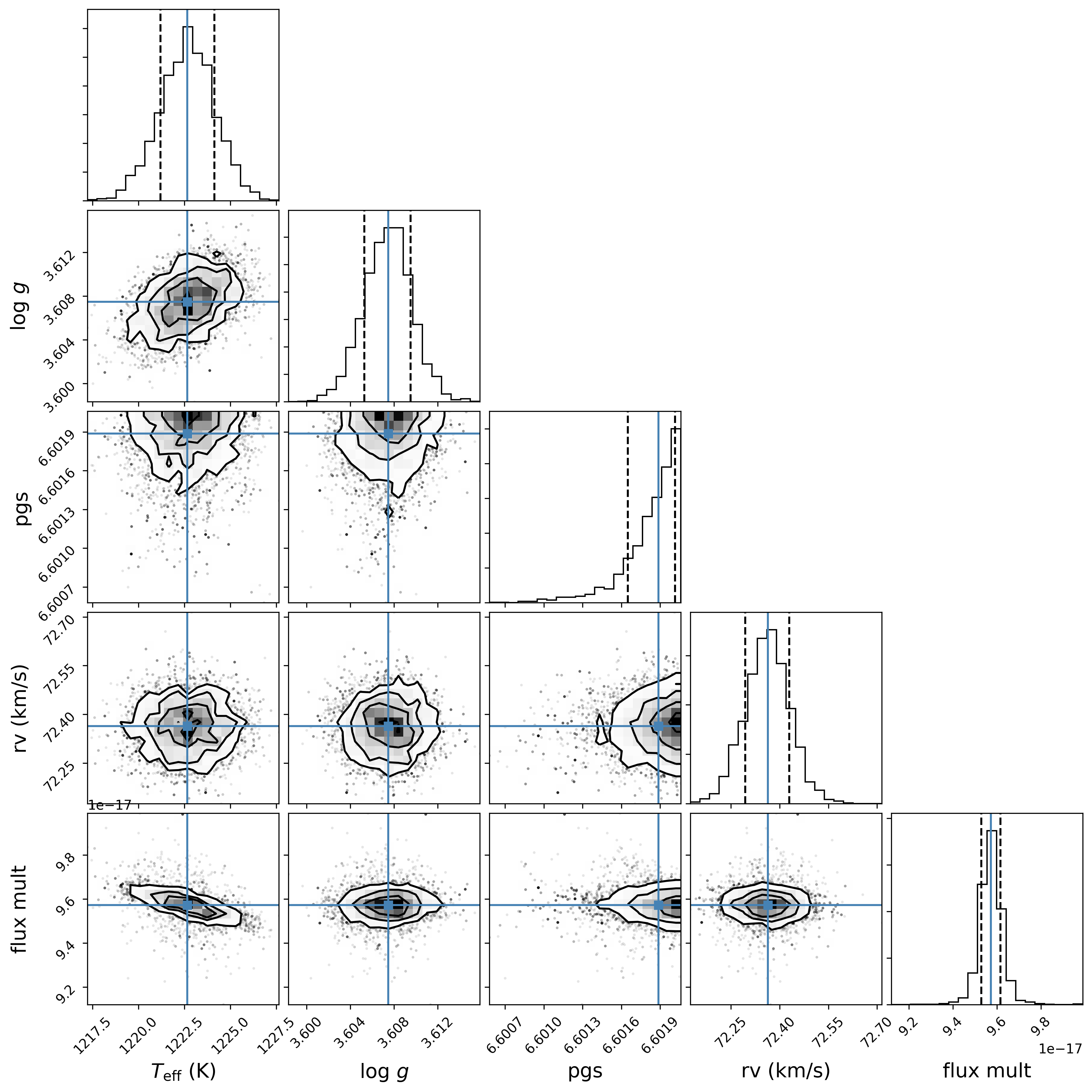}
\caption{Corner plot from our MCMC fits for our continuum subtracted OSIRIS $K$-band spectra. The diagonal shows the marginalized posteriors. The covariances between all the parameters are in the corresponding 2-d histograms. The blue lines represent the 50 percentile, and the dotted lines represent the 16 and 84 percentiles. The "flux mult" corresponds to the dilution factor that scales the model by \((radius)^2 (distance)^{-2}\). Our fit prefers a value near the edge of the grid for pgs, or, $P_{cloud}$.}
\label{fig:corner}
\end{figure*}

We ran our MCMC fitting procedure on both the flux calibrated spectrum and the continuum-subtracted spectrum of VHS 1256b. Objects like VHS 1256b that straddle the L/T transition are notoriously difficult to fit with an unconstrained MCMC, primarily due to the complexity of modeling the clouds at low surface gravities (e.g., \citealt{barman2011}). Therefore, we constrained the bounds of the fits via two methods. First, we used previously derived literature values for temperature and surface gravity as a starting point for a plausible range of values (e.g., \citealt{gauza2015,miles2018}). Then, we plotted the models in those temperature ranges over the data to verify by-eye that the matches were reasonable before running the full MCMC. We therefore narrowed the prior ranges for the temperature to be between 1000-1300 K and a $\log g$ between 3.0-3.5. The best-fit parameters for our flux calibrated data are $T_\mathrm{eff} = 1240^{+1}_{-2}$ K, $\log g = 3.25^{+0.01}_{-0.1}$, and $\log P_{cloud}$ = $6 \pm 0.01$. For the radius, which comes from the multiplicative flux parameter, we found R = 1.00 $\pm$ 0.02 R$_{Jup}$. For our continuum-subtracted data the best-fit parameters were $T_\mathrm{eff} = 1223^{+1}_{-2} 1$~K, $\log g = 3.6^{+0.02}_{-0.01}$, and $\log P_{cloud}$ = $6.6^{+0.07}_{-0.05}$. Radii cannot be derived for the continuum-subtracted data. Figures \ref{fig:continuum_mod} through \ref{fig:corner} show the best-fit spectra overplotted on our data, the residuals from the fits, and the resulting corner plot from our MCMC analysis for the continuum-subtracted spectra.

In order to verify that the temperature estimates we derived from the OSIRIS spectra are robust, we ran our MCMC fitting code using the three custom \textit{PHOENIX} grids on the \cite{gauza2015} optical, $J$-band, $H$-band, and $K$-band spectra and the \cite{miles2018} $L$-band spectra shown in Table 2. We adjusted our MCMC parameters for the \cite{gauza2015} spectra by setting the line-spread function (LSF) to 229.77 km s$^{-1}$, which we derived from the data. We fit each band in the data set separately and obtained $T_\mathrm{eff} = 1201^{+2}_{-1}$ K, $\log g = 4.81 \pm 0.06$, and $\log P_{cloud}$ = $6.59 \pm 0.01$ for the optical+$J$ band spectra, $T_\mathrm{eff} = 1281^{12}_{-23}$ K, $\log g = 3.67^{+0.05}_{-0.04}$, and $\log P_{cloud}$ = $6.06 \pm 0.02$ for the $K$-band spectra, and $T_\mathrm{eff} = 1201^{+2}_{-1}$ K, $\log g = 3.51 \pm 0.01$, and $\log P_{cloud}$ = $5.97^{+0.02}_{-0.03}$ for the $H$-band spectra using our custom \textit{PHOENIX-ICM} grid. We also adjusted our MCMC parameters for the \cite{miles2018} by using an LSF of 106.95 km s$^{-1}$ to account for the lower resolution of the data and obtained an $T_\mathrm{eff} = 1205^{+7}_{-4}$ K, $\log g = 3.53^{+0.03}_{-0.02}$, and $\log P_{cloud}$ = $6.23 \pm 0.02$. 

Since our subsequent analysis of the chemical abundances of VHS 1256 b relies on knowledge of the temperature and gravity, we did additional modeling to look at the comparison between our custom cloud model grid and similar grids with different treatment of clouds. In addition to differences in cloud parameters, each grid of models incorporates different assumptions that lead to systematic differences in the output spectra for the same parameters such as temperature and gravity (e.g., \citealt{oreshenko2020}). These systematics are not captured in the uncertainties that come from each MCMC run output, and therefore those uncertainties, while reported here for completeness, are underestimates of the true uncertainties.  These systematics also partially explain why fitting different parts of the spectra give slightly different best fit values, a very common occurrence for substellar object fits (e.g.,\citealt{cruz2018}). We attempt to account for these systematics by looking at the range of values given from the three \textit{PHOENIX} grids.

We incorporated our \textit{PHOENIX-ACES} grid from \cite{wilcomb2020} and the \textit{PHOENIX-HR8799} grid from \cite{ruffio2021} into our MCMC analysis code, and fit our OSIRIS spectrum using the same procedure described above. The best-fit using \textit{PHOENIX-ACES} yielded $T_\mathrm{eff} = 1500 \pm 5$ K and $\log g = 3.92^{+0.06}_{-0.08}$. The \textit{PHOENIX-HR8799} models generally provided poor matches to the higher resolution data, yielding much higher residuals than other grids.  With this grid, we found $T_\mathrm{eff} = 820^{+2}_{-2}$ K and $\log g = 3.01^{+0.02}_{-0.01}$ as best-fit parameters. We found no fits with \textit{PHOENIX-HR8799} that properly captured the continuum shape of our OSIRIS data. The fits to the continuum were generally poor across all grids, including the \textit{PHOENIX-ICM} grid, and we speculated that this was due to our assumption about grain size in the models (see Section \ref{sec:gs}). We also fit the three \cite{gauza2015} spectra and the \cite{miles2018} spectra with these additional grids. The \textit{PHOENIX-HR8799} grid did a poor job of fitting the \cite{miles2018} spectra, while the \textit{PHOENIX-ACES} grid fit the data well but gave unphysical parameters when the radii were derived shown in Table \ref{tab:atm_param}. Overall, the \textit{PHOENIX-ICM} grid fit the spectra of this object the best and we adopt the values from these fits in this paper. 

Table \ref{tab:fits} shows the results for all atmospheric parameters derived in this paper. We use the range of best-fit values from the OSIRIS continuum-included data to define the adopted parameters for temperature, gravity, and cloud parameter, as the resolved line information coupled with the continuum offers the most constraints on those parameters. To derive more robust uncertainties on the atmospheric parameters that account for the systematics described above, we consider the range of values from all fits using the \textit{PHOENIX-ICM} grid when deriving th allowed values for VHS 1256b. We adopt values of $T_\mathrm{eff} = 1240$~K, with a range of 1200--1300~K, $\log g=3.25$, with a range of 3.25--3.75, and $\log P_{cloud}$ = $6.0$, with a range of 6.0--6.6. For radius, we use the median value from the OSIRIS continuum-included data and the \cite{gauza2015} spectra to arrive at $R=1.0~R_{Jup}$, with a range of 1.0--1.2~$R_{Jup}$. This yields an implied bolometric luminosity of log(L/L$_{\odot}$) = $4.67$, with a range of $4.60$ to $4.70$. These values are all in excellent agreement with previous studies \citep{gauza2015,miles2018}.

\subsubsection{Cloud Grain Size}\label{sec:gs}
VHS 1256 b is an L/T transition object, and these objects spectra are strongly impacted by the depth of clouds and the grain size in the atmosphere \citep{brock2021}. Since our custom \textit{PHOENIX-ICM} grid also varied in grain size, we explored different mean grain sizes to get a better fit to the continuum in our OSIRIS data, as well as a better match to data in all wavebands. We held all other parameters constant, except for the multiplicative factor that is a proxy for radius, and allowed the mean grain size to vary from 0.5 $\mu$m to 3 $\mu$m. For our OSIRIS data, the best fit grain size was about 3$^{+0}_{0.1} \mu$m, which did hit the edge of our grid. For the \cite{gauza2015} spectra, the best fit grain size was 0.51$^{+0.02}_{-0.0094} \mu$m at the short wavelengths, 0.50$^{+0.010}_{-0.0048} \mu$m for the $H$-band spectrum, and 0.99$^{+0.018}_{-0.032}$ for the $K$-band spectrum. The best fit grain size for the \cite{miles2018} spectrum was 0.51$^{+0.11}_{-0.01} \mu$m.

Since the fits to the short and long wavelength data yielded different preferred grain sizes than the OSIRIS data, we examined both 0.5 $\mu$m and 3.0 $\mu$m models across the full range of available wavelengths simultaneously. Figure \ref{fig:extracted_spec} shows all available spectral and photometric data for VHS 1256 b. Overplotted are the two \textit{PHOENIX-ICM} models with two different grain sizes, 0.5 $\mu$m and 3 $\mu$m. The models are scaled to match the continuum flux at $K$-band, which in turn is derived using the $K$-band magnitude in \citet[$K_s = 14.665 \pm 0.01$;][]{gauza2015}.  First, we note that the lower resolution $K$ band spectra from \citet{gauza2015} spectra are a good overall match to the OSIRIS data.  We note that when we look at the model in comparison to the flux calibrated shorter wavelength data, the 3 $\mu$m grain size model is a good match overall though not a perfect match to the continuum shape.  However, the 0.5 $\mu$m model, the model strongly underpredicts the short wavelength flux if it is scaled to match $K$ band.  This means that the best-fit of 0.5 $\mu$m at these short wavelengths was dependent on scaling the radius to much larger values was required for the $K$ band.  Indeed, with a the best-fit value for $K$ band is a radius of $\sim$1 R$_{Jup}$, whereas the best fit at short wavelengths with a small grain size is $\sim$1.4 R$_{Jup}$.  This implies that the fits were being pulled by the continuum shape when performed individually, and a comprehensive look at data at all wavelengths is more appropriate for determining the grain size.  We do note that in the case of the long wavelength data from  \cite{miles2018} spectra and the photometry from \cite{rich2016}, the red side of the spectrum matches the 0.5 $\mu$m model better, and the radius is consistent with $K$ band and roughly the same for either grain size.  However, the overall shape of the \cite{miles2018} spectrum does not match either model. Taken together, we conclude that a larger median grain size of 3.0 $\mu$m is appropriate for this object.  

\subsection{Mole Fractions of CO, H$_2$O, and CH$_4$}
With best fit values for temperature, surface gravity, and clouds, we can derive abundances of CO, H$_2$O, and CH$_4$ using our OSIRIS $K$-band spectra. Once best fit values were determined for $T_\mathrm{eff}$, $\log g$, and $\log P_{cloud}$, we fixed those parameters to generate a grid of spectra with scaled mole fractions of these molecules, which are visible in the $K$-band \citep{barman2015}. We used solar metallicity for this grid because there is no evidence for non-solar metallicity. Therefore, the unscaled mixing ratios in this grid will match the composition and mole fractions of the Sun. The molecular abundances of CO, CH$_4$, and H$_2$O were scaled relative to their initial values from 0 to 1000 using a uniform logarithmic sampling, resulting in 25 synthetic spectra. We fit for the mole fraction of H$_2$O first, holding CO and CH$_4$ at their initial values. Next, the H$_2$O mole fraction was set to its nominal value, and we fit for scaled CO.  We did the same analysis for CH$_4$.

\begin{figure*}
\epsscale{0.95}
\plotone{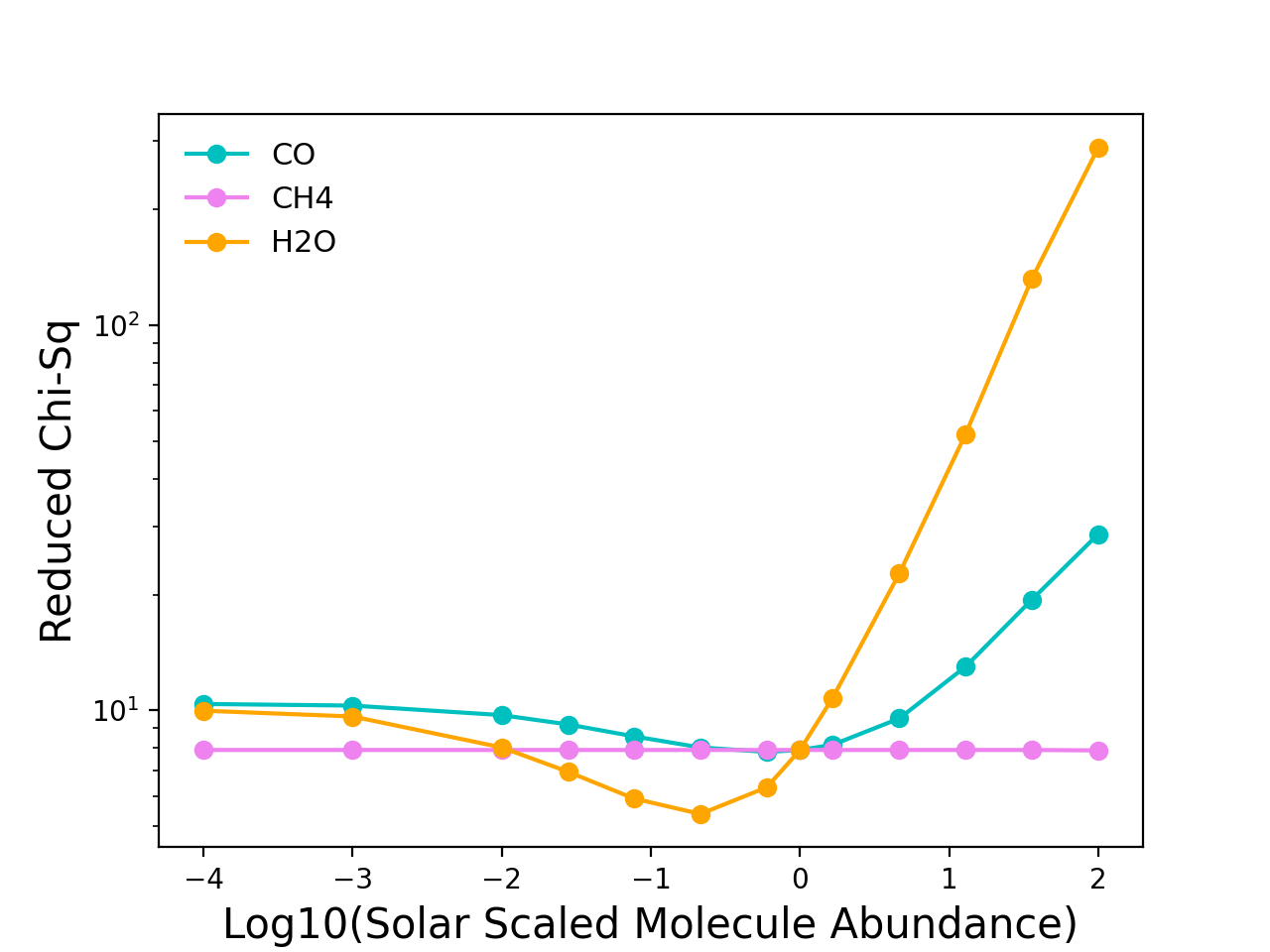}
\caption{Results of $T_\mathrm{eff}$ = 1240 K, $\log g$ = 3.25, $\log P_{cloud}$ = 6, and grain size = 3 $\mu$m model fits with varying mole fractions for both H$_2$O, CO, and CH$_4$ to our continuum-subtracted OSIRIS spectrum. The mole fractions are given in units relative to the ratio in the Sun, such that a value of zero implies the solar value.  The scalings of CO prefer solar/subsolar and the scalings of H$_2$O prefer values subsolar. From these fits we find C/O = 0.869$_{-0.24}^{+0.09}$.} 
\label{fig:chisq_co_h2o}
\end{figure*}

Figure \ref{fig:chisq_co_h2o} shows the resulting \(\chi^2\) distribution as a function of CO, H$_2$O, and CH$_4$ mole fraction. The models with the lowest \(\chi^2\) when compared to the flattened data gave us the best-fits for both H$_2$O and CO.  For CH$_{4}$, there is no change in  \(\chi^2\) with any change in the mole fraction, suggesting that if any methane is present in the spectrum, it is below the detection threshold of our data.  We exclude CH$_{4}$ from further analysis. The best fit for H$_2$O had a scaling of 0.215$^{+0.784}_{-0.187}$, and the best fit for CO had a scaling of 0.599$^{+1.061}_{-0.384}$.  To calculate the 1-$\sigma$ uncertainties in each mole fraction value, we used the values from models within $\pm$1 of our lowest \(\chi^2\). 

\begin{figure*}
\epsscale{0.95}
\plotone{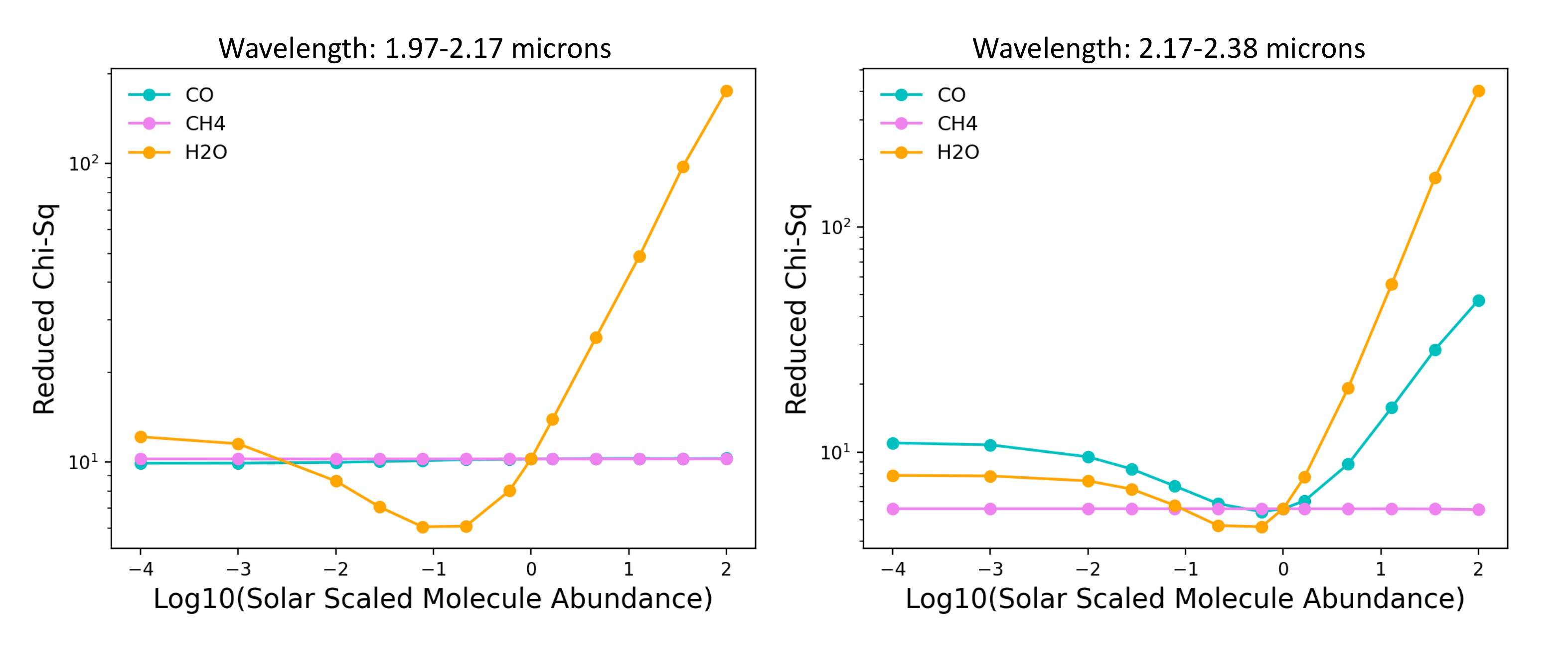}
\caption{\textbf{Left:} The results of our \(\chi^2\) analysis on the first half of our continuum subtracted OSIRIS data. The CO abundances vary minimally during this analysis because of the lack of CO lines in the first half of our spectra. The best-fit water is still subsolar, with the lowest \(\chi^2\) values not including solar. \textbf{Right:} The results of our \(\chi^2\) analysis on the second half of our continuum subtracted OSIRIS data. The values allowed by the \(\chi^2\) analysis include the solar abundance model, with the best-fit H$_2$O molecular abundance being slightly subsolar.}
\label{fig:1st_2nd_half}
\end{figure*}

Because of the potential issues with telluric correction in the blue side of the spectrum mentioned in Section \ref{sec:data}, we explored the mole fractions on different portions of the spectra, splitting it in half and performing the fit again.  Shown in the left panel of Figure \ref{fig:1st_2nd_half}, for the first half of the spectrum, the best fit H$_2$O mole fraction is 0.215 and CO is unchanging as there are no CO features in this section of the spectra. For the second half of the OSIRIS spectra, shown in right panel of Figure \ref{fig:1st_2nd_half}, the H$_2$O scaled mole fraction is higher at 0.599 and the uncertainties encompass the lowest \(\chi^2\) from the full spectrum analysis. The lowest \(\chi^2\) CO matches the values found when fitting the full spectrum (Figure \ref{fig:chisq_co_h2o}). Because of concerns about the blue half of the spectrum, we elect to use the C and O values from the second half of the spectrum.  This yields an H$_2$O mole fraction of 0.599$^{+0.401}_{-0.384}$ and a CO mole fraction of 0.599$^{+1.061}_{-0.384}$.

\subsection{C/O Ratios}

In our previous work, we have used OSIRIS data to constrain the C/O ratios of directly imaged companions to massive stars such as HR 8799 b, c, and d, and $\kappa$ Andromedae b (e.g., \citealt{konopacky2013,wilcomb2020,ruffio2021}). For giant planets formed by rapid gravitational instabilities, their atmospheres should have elemental abundances that are the same as their host stars \citep{helled2009}. If giant planets form by a multi-step core accretion process, it has been suggested that there could be a range of elemental abundances possible \citep{oberg2011,madhu2019}. In this scenario, the abundances of giant planets' atmospheres formed by core/pebble accretion are highly dependent on the location of formation relative to CO, CO$_2$, and H$_2$O frost lines and the amount of solids acquired by the planet during runaway accretion phase.  This can be diagnosed using the C/O ratio. 

For VHS 1256 b, a substellar companion to a small binary pair, it is unclear if C/O is a useful formation diagnostic, as it is likely to have formed like a binary star rather than a planet (see Section \ref{sec:discussion}).  However, we still compute the ratio here and discuss the implications below.

The C/O ratio dependence on atmospheric mole fractions (N) is

\[\frac{C}{O}=\frac{N(CH_4)+N(CO)}{N(H_2O)+N(CO)},\]

\noindent and for small amounts of CH$_{4}$, as in VHS 1256 b's case, the C/O ratio can be determined by H$_2$O and CO alone \citep{barman2015}. The C/O ratio we derive for VHS 1256 b is 0.590$_{-0.384}^{+0.280}$. The large uncertainties encompass potential systematics and any differences in C/O ratio from the two halves of the spectrum. Variability of the object is unlikely to affect the measurement of C/O because CO is formed higher in the atmosphere than the predicted location of cloud decks.  The variability at K band has been shown to be less significant than at other wavelengths (e.g., \citealt{vos2019,bowler2020}). Figure \ref{fig:panel_c/o} shows our lowest \(\chi^2\) model in the center panel and two extreme cases to illustrate the sensitivity our OSIRIS moderate resolution spectra has to the C/O ratio.

\section{Discussion and Conclusions}\label{sec:discussion}
Moderate resolution spectroscopy has expanded our knowledge of young directly imaged planet atmospheres. VHS 1256 b provides an interesting diagnostic because of the low contrast between the companion and the host system, and the thick cloudy atmosphere. Our results of $T_\mathrm{eff} = 1240$~K, with a range of 1200--1300~K, and $\log g=3.25$, with a range of 3.25--3.75 agree with the results from \cite{gauza2015} that posit VHS 1256 b is a low surface gravity object and is likely a planetary mass object in a youthful system. Our effective temperature is also in agreement with the literature regardless of the new distance measured by Gaia EDR3, 22.2$^{+1.1}_{-1.2}$ pc, which changed the distance from 12.7$\pm$1.0 pc \citep{gauza2015}. \cite{dupuy2020} also found a distance  22.22 $\pm$ 1.19 pc from CFHT parallax measurements. We did not detect methane in our $K$-band spectra, which is in agreement with very weak methane absorption in the $L$-band spectra pointing towards non-equilibrium chemistry from \cite{miles2018}. 

Our OSIRIS data on VHS 1256 b, coupled with data from the literature, preferred a thick-cloud model, $\log P_{cloud}$ = $6.0$ with a range of 6.0--6.6, with a 3 $\mu$m mean grain size. Recent works have demonstrated that earlier L-type objects have higher-altitude clouds with smaller grains (0.25--0.50 $\mu$m), and later T-type objects exhibit deeper clouds and larger grains ($\ge$ 1 $\mu$m; \citealt{brock2021}). We conclude that VHS 1256 b has deep clouds with a large mean grain size that is consistent with being near or just beyond the L/T transition, as indicated by its L7 spectral type. 

Cloudy atmospheres of these Jovian-like objects are not well understood. Data at longer wavelengths can more deeply probe these cloud parameters and can inform the models to better understand the processes occurring in cloudy giant planet atmospheres. VHS 1256 b is also extremely variable with near infrared photometry in 2MASS J, H, and K/K$_s$ that varies by $\sim$37, $\sim$13, and $\sim$3 percent respectively \citep{miles2018}. This can lead to different colors and flux calibrations that can influence interpreted atmospheric parameters. Understanding the variability of this object is important for further constraints on effective temperature and other atmospheric parameters. Our derivation of cloud properties, including a notional grain size distribution, will inform studies of variability of VHS 1256 b. 

\begin{figure*}
\epsscale{0.95}
\plotone{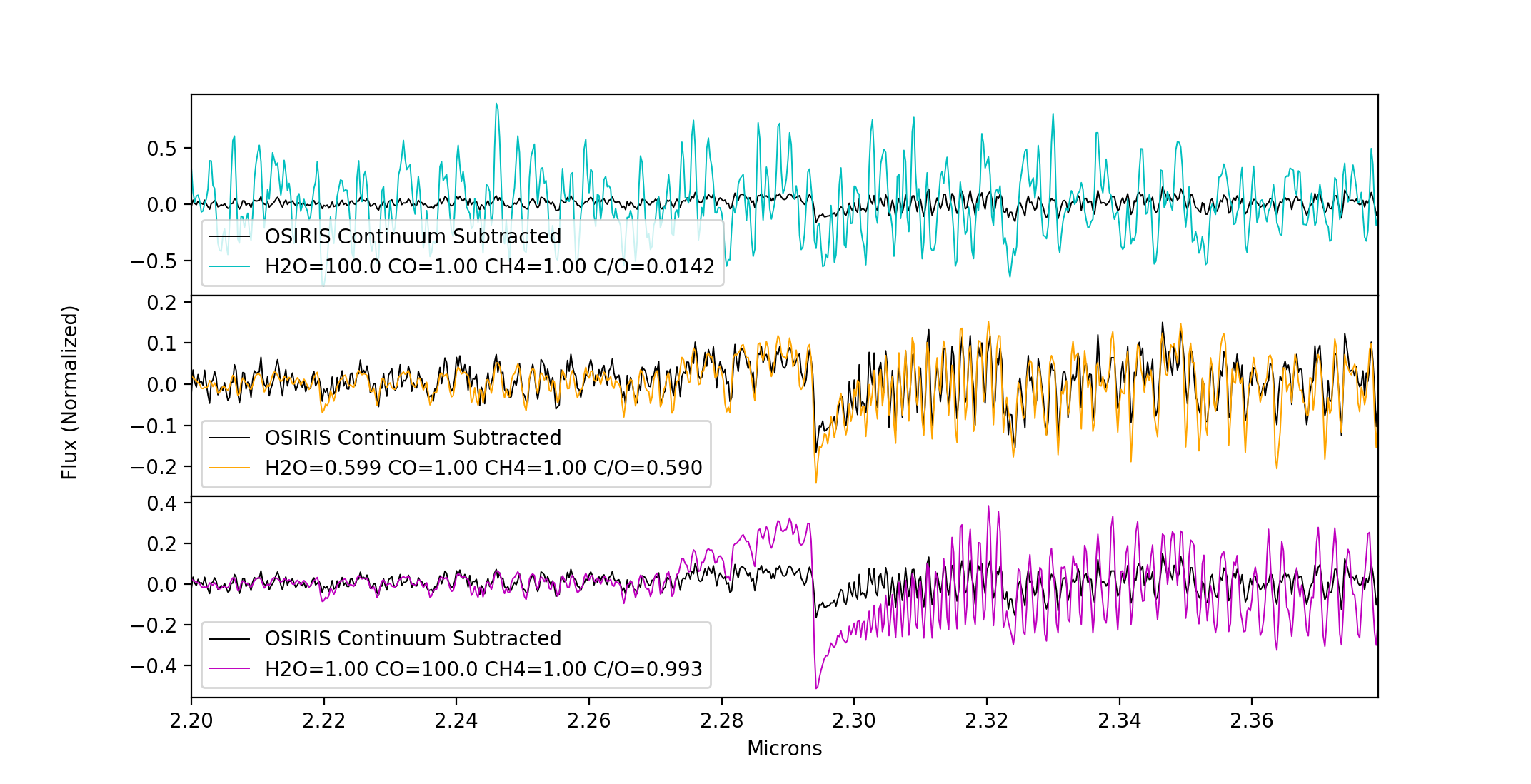}
\caption{Our fully reduced, combined, and continuum subtracted moderate-resolution OSIRIS $K$-band spectra of VHS 1256 b plotted alongside three models of T=1240 K and $\log g$=3.2 with varying C/O. The top panel shows the super-solar H$_2$O model with C/O=0.0142 in blue. The middle panel shows our best fit model with subsolar H$_2$O and solar CO and CH$_4$, which gives C/O=0.590. The bottom panel shows a supersolar CO model with C/O=0.993. This demonstrates how well moderate resolution spectroscopy can constrain C/O ratios of planetary mass objects.} 
\label{fig:panel_c/o}
\end{figure*}

Formation of objects in the gas giant planet mass regime is of considerable interest, whether the companion orbits a low or high mass star.  Formation diagnostics such as C/O ratio can provide context and potentially point to a formation pathway, although interpretation of the C/O ratio is complicated by a variety of phenomena (e.g., \citealt{madhu2019}). The mass ratio of the system is 0.16, which is significantly larger than other directly imaged planets such as HR 8799 (q $\sim$ 5 $\times$ $10^{-3}$; \citealt{fabrycky2010}) suggesting that the companion was likely too large to form in a protoplanetary disk around the primary \citep{rich2016}.  In measuring the C/O ratio here for VHS 1256 b, we seek instead to use observations of a similar object to the directly imaged planets to look for systematics in our modeling that could suggest any biases in computation of C/O. We derived a C/O ratio of 0.590$_{-0.384}^{+0.280}$ for VHS 1256 b. Our finding of a roughly solar C/O ratio implies that these systematics are not present, which would be suggested by a ratio that is very different from solar.  An unusual C/O ratio compared to the stellar value in a low mass binary is unexpected, and would suggest issues with modeling L/T transition objects. It is likely that VHS 1256 b formed like a binary star rather than a planet. There are studies that claim metallicity measurements could constrain formation location for these types of objects, but future work is needed to obtain these measurements for the host binary and the wide orbit companion \citep{liufan2021}.  While we do not measure a C/O ratio for the host star because we did not have moderate resolution spectroscopy for it, there is no indication that a non-solar metallicity is required to describe the spectra in hand \citep{gauza2015}.  Thus, we are confident our methodology is sound and reliably provides C/O ratios for directly imaged companions.  Indeed, Figure \ref{fig:panel_c/o} shows that very large or small C/O ratios are easily probed and ruled out by moderate-to-high resolution data, and are thus more trustworthy than values computed on low resolution spectra.

VHS 1256 b now represents an eighth case of an imaged planetary mass object, in addition to the four HR 8799 planets \citep{konopacky2013,barman2015,molliere2020,ruffio2021}, $\kappa$ And b \citep{wilcomb2020}, HIP 65426 b \citep{petrus2021}, and TYC 8998-760-1 b (YSES-1 b \citealt{zhang2021}), where the C/O ratio formation diagnostic did not reveal ratios that clearly point to formation via core/pebble accretion. The scenario cannot be completely ruled out given the uncertainties in the data and the range of possible C/O ratios predicted by models (e.g., \citealt{madhu2019}). Therefore, other probes of formation such as the $^{12}$CO/$^{13}$CO isotope ratio \citep{zhang2021} and metallicity, will be needed to shed more light on this intriguing population of giant companions. 

VHS 1256 b has already been targeted with a higher resolution instrument, NIRSPEC at the Keck II 10 m telescope, which allowed for measurements of radial velocity and rotational velocity \citep{bryan2018}. The next steps for the VHS 1256 system will be space-based spectroscopy to verify the C/O ratio and other atmospheric properties. VHS 1256 b is a target for the High Contrast Imaging of Exoplanets and Exoplanetary Systems with JWST Early Release Science (ERS) Program \citep{hinkley2017}. VHS 1256 b will be studied using NIRSpec and MIRI to analyze the atmosphere at longer wavelengths to constrain non-equilibirum chemistry, C/O ratio, and variability of the object. We can also determine whether the bulk population of directly imaged planets show C/O ratios consistent with solar/stellar values by continuing to obtain moderate- or high-resolution spectra of these wide orbit, self-luminous companions. If the population of these objects show distinct C/O from closer in transit giant planets, this could point towards differing formation pathways for these companions. 

\begin{acknowledgments}

The authors thank observing assistant John Pelletier and support astronomer Jim Lyke for their help obtaining these observations.  K.K.W.H., Q.M.K, T.S.B, and L.S.B. acknowledge support by the National Aeronautics and Space Administration under Grants/Contracts/Agreements No.NNX17AB63G and  80NSSC21K0573 issued through the Astrophysics Division of the Science Mission Directorate.  T.S.B. also acknowledges support by the National Science Foundation under Grant No. 1614492. Any opinions, findings, and conclusions or recommendations expressed in this paper are those of the author(s) and do not necessarily reflect the views of the National Aeronautics and Space Administration. The data presented herein were obtained at the W. M. Keck Observatory, which is operated as a scientific partnership among the California Institute of Technology, the University of California and the National Aeronautics and Space Administration. The Observatory was made possible by the generous financial support of the W. M. Keck Foundation. The authors wish to recognize and acknowledge the very significant cultural role and reverence that the summit of Maunakea has always had within the indigenous Hawaiian community. We are most fortunate to have the opportunity to conduct observations from this mountain.

Portions of this work were conducted at the University of California, San Diego, which was built on the unceded territory of the Kumeyaay Nation, whose people continue to maintain their political sovereignty and cultural traditions as vital members of the San Diego community. 

\end{acknowledgments}

\facilities{Keck/OSIRIS}

\software{\textit{emcee} \citep{foreman-mackey2013}, \textit{SMART} \citep{hsu2021_smart,hsu2021}}

\begin{deluxetable}{lccc} 
\tabletypesize{\scriptsize} 
\tablewidth{0pt}
\tablecaption{Summary of VHS 1256 AB b System \label{tab:sum_system}} 
\tablehead{ 
 \colhead{Parameter} & \colhead{Central Binary} & \colhead{Companion} & \colhead{Reference}}
\startdata 
Distance (pc) & 22.2$^{+1.1}_{-1.2}$ &  - & \tablenotemark{a} \\
Separation (arcsec) & $0.109 \pm 0.0018$ &  $8.06 \pm 0.03$ & \tablenotemark{b,c} \\
Projected Separation (au) & $2.42 \pm 0.0399$ &  $102 \pm 9$  & \tablenotemark{b,c}\\
Spectral Type & M7.5$ \pm 0.5$ & L7.0$ \pm 1.5$ & \tablenotemark{c}\\
\hline
\multicolumn{3}{c}{Apparent Magnitudes} \\
\hline
2MASS $J$-band & 11.018$ \pm 0.023$ & 16.662$ \pm 0.287$ & \tablenotemark{d} \\
2MASS $H$-band & 10.473$ \pm 0.023$ & 15.595$ \pm 0.209$ & \tablenotemark{d} \\
2MASS $K_s$-band & 10.044$ \pm 0.021$ & 14.568$ \pm 0.121$ & \tablenotemark{d} \\
VHS $Y$-band & $<11.72 $ & 18.558$ \pm 0.051$ & \tablenotemark{e} \\
VHS $J$-band & $<11.36 $ & 17.136$ \pm 0.020$ & \tablenotemark{e} \\
VHS $H$-band & $<11.02 $ & 15.777$ \pm 0.015$ & \tablenotemark{e} \\
VHS $K_s$-band & $<10.42 $ & 14.665$ \pm 0.010$ & \tablenotemark{e} \\
Subaru/IRCS $L$-band & 9.76$ \pm 0.03$ & 12.99$ \pm 0.04$ & \tablenotemark{f} \\
\enddata
\tablerefs{(a) \cite{dupuy2020}, (b)  \cite{stone2016}, (c)  \cite{gauza2015}, (d) \cite{skrutskie2006}, (e) \cite{mcmahon2013},  (f) \cite{rich2016}}
\end{deluxetable}


\begin{deluxetable*}{lccc} 
\tabletypesize{\scriptsize} 
\tablewidth{0pt} 
\tablecaption{Summary of atmospheric parameters derived from MCMC fits for the Primary.\label{tab:primary}}
\label{tab:atm_param_prim}
\tablehead{ 
  \colhead{Spectra} & \colhead{Effective Temperature} & \colhead{Surface Gravity} &
  \colhead{Metallicity}\\
  \colhead{VHS 1256 b} & \colhead{${T}_\mathrm{{eff}}$ (K)} & \colhead{$\log g$} & \colhead{$M/H$} 
}
\startdata 
\hline
\multicolumn{4}{c}{Göttingen PHOENIX Spectral Library} \\
\hline
Gauza Primary $K$-band & $2756^{+13}_{-15}$ & $4.99^{+0.0076}_{-0.015}$ & $0.0093^{+0.012}_{-0.0063}$ \\
Gauza Primary Optical+$J$ band & $2668^{+8}_{-7}$ & $3.57 \pm 0.060$ & $0.20 \pm 0.040$ \\
Gauza Primary $H$-band & $2758^{+15}_{-18}$ & $3.92^{+0.090}_{-0.40}$ & $0.39^{+0.14}_{-0.35}$ \\
\hline
Allowed Range of Values & 2661 - 2773 & 3.51 - 5.00 & 0.0030 - 0.53  \\
\enddata
\end{deluxetable*}

\begin{deluxetable*}{lcccccc} 
\tabletypesize{\scriptsize} 
\tablewidth{0pt} 
\tablecaption{Summary of atmospheric parameters derived from MCMC fits.\label{tab:fits}}
\label{tab:atm_param}
\tablehead{ 
  \colhead{Spectra} & \colhead{Effective Temperature} & \colhead{Surface Gravity} &
  \colhead{Cloud Layer} & \colhead{Grain size} & \colhead{Radius} & \colhead{Luminosity}\\
  \colhead{VHS 1256 b} & \colhead{${T}_\mathrm{{eff}}$ (K)} & \colhead{$\log g$} & \colhead{$\log P_{cloud}$}
  & \colhead{$\mu$m} &\colhead{($R_\mathrm{Jup}$)} &\colhead{$\log_{10}\left(\frac{L}{L_\odot}\right)$} 
}
\startdata 
\multicolumn{6}{c}{PHOENIX-ACES} \\
\hline
OSIRIS Including Continuum & $1500^{+5}_{-5}$ & $3.92^{+0.06}_{-0.08}$ & n/a & n/a & $0.64$ & $-4.73$ \\
OSIRIS Continuum Subtracted & $1500^{+5}_{-5}$ & $4.47^{+0.01}_{-0.02}$ & n/a & n/a & n/a & n/a \\
Gauza $K$-band & $1839^{+6}_{-7}$ & $4.83^{+0.01}_{-0.02}$ & n/a & n/a & $0.41$ & $-4.76$ \\
Gauza Optical+$J$ band & $1659^{+3}_{-3}$ & $3.72^{+0.01}_{-0.01}$ & n/a & n/a & $0.52$ & $-4.73$ \\
Gauza $H$-band & $1658^{+2}_{-2}$ & $4.65^{+0.02}_{-0.02}$ & n/a & n/a & $0.54$ & $-4.70$ \\
Miles $L$-band & $1096^{+12}_{-13}$ & $4.38^{+0.11}_{-0.13}$ & n/a & n/a & $1.18$ & $-4.74$ \\
\hline
\multicolumn{6}{c}{PHOENIX-ICM} \\
\hline
OSIRIS Including Continuum & $1240^{+1}_{-2}$ & $3.25^{+0.01}_{-0.1}$ & $6.0 \pm 0.01$ & $3.0 \pm 0.1$ & $1.0$ & $-4.67$ \\
OSIRIS Continuum Subtracted & $1223^{+1}_{-2}$ & $3.6^{+0.02}_{-0.01}$ & $6.6^{+0.07}_{-0.05}$ & $2.9^{+0.001}_{-0.01}$ & n/a & n/a \\
Gauza $K$-band & $1281^{+12}_{-23}$ & $3.67^{+0.05}_{-0.04}$ & $6.06 \pm 0.02$ & $0.99^{+0.018}_{-0.032}$ & $0.92$ & $-4.69$ \\
Gauza Optical+$J$-band & $1201^{+2}_{-1}$ & $4.81^{+0.06}_{-0.06}$ & $ 6.59 \pm 0.01$ & $0.51^{0.02}_{0.0094}$ & $1.2$ & $-4.57$ \\
Gauza $H$-band & $1201^{+2}_{-1}$ & $3.51^{+0.01}_{-0.01}$ & $5.97^{+0.02}_{-0.03}$ & $0.05^{+0.010}_{-0.0048}$ & $1.18$ & $-4.58$ \\
Miles $L$-band & $1205^{+7}_{-4}$ & $3.53^{+0.03}_{-0.02}$ & $6.23 \pm 0.02$ & $0.51^{+0.11}_{-0.01}$ & $1.13$ & $-4.61$ \\
\hline
\multicolumn{6}{c}{PHOENIX-HR8799} \\
\hline
OSIRIS Including Continuum & $820^{+2}_{-2}$ & $3.01^{+0.02}_{-0.01}$ & $5.7^{+0}_{-0}$ & n/a & $2.9$ & $-4.46$ \\
OSIRIS Continuum Subtracted & $1300^{+1}_{-2}$ & $3.0^{+0.02}_{-0.0}$ & $6.0^{+0.01}_{-0.01}$ & n/a & n/a & n/a  \\
Gauza $K$-band & $1298^{+1}_{-3}$ & $3.28^{+0.02}_{-0.02}$ & $5.7^{+0.01}_{-0.0}$ & n/a & $0.9$ & $-4.68$ \\
Gauza Optical+$J$-band & $1097^{+17}_{-21}$ & $3.28^{+0.28}_{-0.16}$ & $5.96^{+0.14}_{-0.13}$ & n/a & $1.8$ & $-4.37$ \\
Gauza $H$-band & $1031^{+9}_{-10}$ & $3.01^{+0.02}_{-0.01}$ & $5.84^{+0.04}_{-0.03}$ & n/a & $1.8$ & $-4.48$ \\
Miles $L$-band & $836^{+7}_{-5}$ & $3.06^{+0.01}_{-0.01}$ & $5.71^{+0.02}_{-0.01}$ & n/a & $2.1$ & $-4.71$ \\
\hline
Adopted Values & 1240 & 3.25 & 6.0 & 3.0 & 1.0 & -4.67 \\
Allowed Range of Values & 1200 - 1300 & 3.25 - 3.75 & 6.0 - 6.6 & 1.0 - 3.0 & 1.0 - 1.2 & -4.60 - -4.70 \\
\enddata
\tablenotetext{-}{Based on our fitting and a comparison of all datasets, we have chosen to adopt the best-fit values from the PHOENIX-ICM fit to the OSIRIS data with the continuum included.  However, we allow for uncertainties that encompass the range of most fits to all datasets using the PHOENIX-ICM grid, excluding the surface gravity value from the \citet{gauza2015} optical data, which is an outlier.}
\end{deluxetable*}

\begin{figure*}
\epsscale{0.95}
\plotone{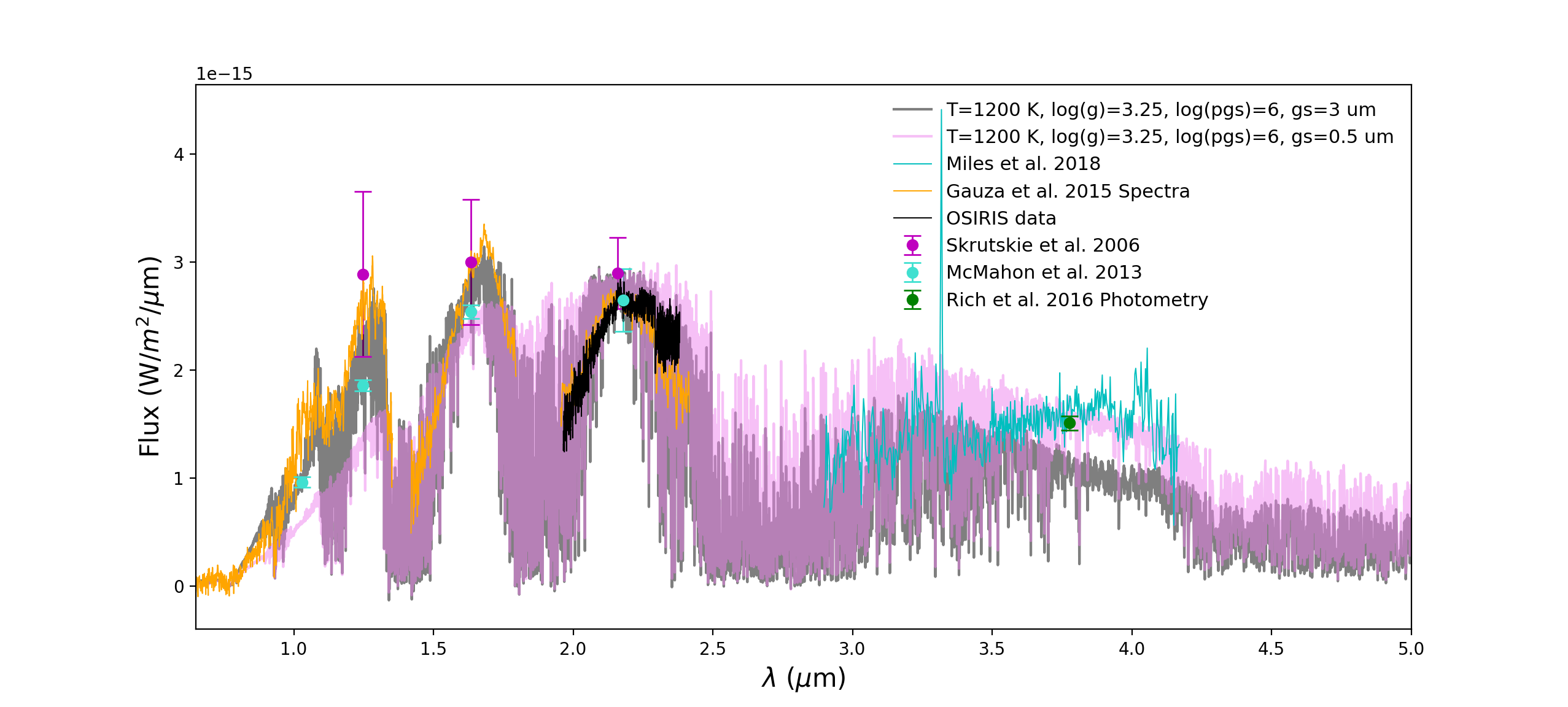}
\caption{Our fully reduced, combined, and flux calibrated moderate-resolution OSIRIS $K$-band spectra of VHS 1256 b in black, plotted alongside our best fit model with two different grain sizes, 3$\mu$m in gray and 0.5$\mu$m in pink, and spectra from \cite{miles2018} in blue, spectra from \cite{gauza2015}, and magnitudes from \cite{rich2016}, \cite{skrutskie2006}, and \cite{mcmahon2013}. The 3 $\mu$m grain size model fits the majority of the data except for the red end of the $L$-band spectra and the \cite{rich2016} photometry point. The 0.5 $\mu$m grain size model underpredicts the flux in the optical, $J$-, and $H$-band, but does fit the $L$-band spectra and photometry point. We adopt the 3 $\mu$m model as the best-fit for the majority of the available data.} 
\label{fig:megaplot}
\end{figure*}

\clearpage
\bibliography{bibliography.bib}

\begin{thebibliography}{}
\expandafter\ifx\csname natexlab\endcsname\relax\def\natexlab#1{#1}\fi
\providecommand{\url}[1]{\href{#1}{#1}}
\providecommand{\dodoi}[1]{doi:~\href{http://doi.org/#1}{\nolinkurl{#1}}}
\providecommand{\doeprint}[1]{\href{http://ascl.net/#1}{\nolinkurl{http://ascl.net/#1}}}
\providecommand{\doarXiv}[1]{\href{https://arxiv.org/abs/#1}{\nolinkurl{https://arxiv.org/abs/#1}}}

\bibitem[{{Allard} {et~al.}(2001){Allard}, {Hauschildt}, {Alexander},
  {Tamanai}, \& {Schweitzer}}]{allard2001}
{Allard}, F., {Hauschildt}, P.~H., {Alexander}, D.~R., {Tamanai}, A., \&
  {Schweitzer}, A. 2001, \apj, 556, 357, \dodoi{10.1086/321547}

\bibitem[{{Bailey} {et~al.}(2013){Bailey}, {Hinz}, {Currie}, {Su}, {Esposito},
  {Hill}, {Hoffmann}, {Jones}, {Kim}, {Leisenring}, {Meyer}, {Murray-Clay},
  {Nelson}, {Pinna}, {Puglisi}, {Rieke}, {Rodigas}, {Skemer}, {Skrutskie},
  {Vaitheeswaran}, \& {Wilson}}]{bailey2013}
{Bailey}, V., {Hinz}, P.~M., {Currie}, T., {et~al.} 2013, \apj, 767, 31,
  \dodoi{10.1088/0004-637X/767/1/31}

\bibitem[{{Barman} {et~al.}(2015){Barman}, {Konopacky}, {Macintosh}, \&
  {Marois}}]{barman2015}
{Barman}, T.~S., {Konopacky}, Q.~M., {Macintosh}, B., \& {Marois}, C. 2015,
  \apj, 804, 61, \dodoi{10.1088/0004-637X/804/1/61}

\bibitem[{{Barman} {et~al.}(2011){Barman}, {Macintosh}, {Konopacky}, \&
  {Marois}}]{barman2011}
{Barman}, T.~S., {Macintosh}, B., {Konopacky}, Q.~M., \& {Marois}, C. 2011,
  \apj, 733, 65, \dodoi{10.1088/0004-637X/733/1/65}

\bibitem[{{Biller} {et~al.}(2013){Biller}, {Liu}, {Wahhaj}, {Nielsen},
  {Hayward}, {Males}, {Skemer}, {Close}, {Chun}, {Ftaclas}, {Clarke}, {Thatte},
  {Shkolnik}, {Reid}, {Hartung}, {Boss}, {Lin}, {Alencar}, {de Gouveia Dal
  Pino}, {Gregorio-Hetem}, \& {Toomey}}]{biller2013}
{Biller}, B.~A., {Liu}, M.~C., {Wahhaj}, Z., {et~al.} 2013, \apj, 777, 160,
  \dodoi{10.1088/0004-637X/777/2/160}

\bibitem[{{Blake} {et~al.}(2010){Blake}, {Charbonneau}, \& {White}}]{blake2010}
{Blake}, C.~H., {Charbonneau}, D., \& {White}, R.~J. 2010, \apj, 723, 684,
  \dodoi{10.1088/0004-637X/723/1/684}

\bibitem[{{Bowler}(2016)}]{bowler2016}
{Bowler}, B.~P. 2016, \pasp, 128, 102001,
  \dodoi{10.1088/1538-3873/128/968/102001}

\bibitem[{{Bowler} {et~al.}(2020){Bowler}, {Zhou}, {Morley}, {Kataria},
  {Bryan}, {Benneke}, \& {Batygin}}]{bowler2020}
{Bowler}, B.~P., {Zhou}, Y., {Morley}, C.~V., {et~al.} 2020, \apjl, 893, L30,
  \dodoi{10.3847/2041-8213/ab8197}

\bibitem[{{Brock} {et~al.}(2021){Brock}, {Barman}, {Konopacky}, \&
  {Stone}}]{brock2021}
{Brock}, L., {Barman}, T., {Konopacky}, Q.~M., \& {Stone}, J.~M. 2021, \apj,
  914, 124, \dodoi{10.3847/1538-4357/abfc46}

\bibitem[{{Bryan} {et~al.}(2018){Bryan}, {Benneke}, {Knutson}, {Batygin}, \&
  {Bowler}}]{bryan2018}
{Bryan}, M.~L., {Benneke}, B., {Knutson}, H.~A., {Batygin}, K., \& {Bowler},
  B.~P. 2018, Nature Astronomy, 2, 138, \dodoi{10.1038/s41550-017-0325-8}

\bibitem[{{Burgasser} {et~al.}(2016){Burgasser}, {Lopez}, {Mamajek},
  {Gagn{\'e}}, {Faherty}, {Tallis}, {Choban}, {Tamiya}, {Escala}, \&
  {Aganze}}]{burgasser2016}
{Burgasser}, A.~J., {Lopez}, M.~A., {Mamajek}, E.~E., {et~al.} 2016, \apj, 820,
  32, \dodoi{10.3847/0004-637X/820/1/32}

\bibitem[{{Chin} {et~al.}(2012){Chin}, {Wizinowich}, {Campbell}, {Chock},
  {Cooper}, {James}, {Lyke}, {Mastromarino}, {Martin}, {Medeiros}, {Morrison},
  {Neyman}, {Panteleev}, {Stalcup}, {Tucker}, {Wetherell}, \& {van
  Dam}}]{chin2012}
{Chin}, J. C.~Y., {Wizinowich}, P., {Campbell}, R., {et~al.} 2012, in Society
  of Photo-Optical Instrumentation Engineers (SPIE) Conference Series, Vol.
  8447, Adaptive Optics Systems III, ed. B.~L. {Ellerbroek}, E.~{Marchetti}, \&
  J.-P. {V{\'e}ran}, 84474F, \dodoi{10.1117/12.925807}

\bibitem[{{Cruz} {et~al.}(2018){Cruz}, {N{\'u}{\~n}ez}, {Burgasser},
  {Abrahams}, {Rice}, {Reid}, \& {Looper}}]{cruz2018}
{Cruz}, K.~L., {N{\'u}{\~n}ez}, A., {Burgasser}, A.~J., {et~al.} 2018, \aj,
  155, 34, \dodoi{10.3847/1538-3881/aa9d8a}

\bibitem[{{Dupuy} {et~al.}(2020){Dupuy}, {Liu}, {Magnier}, {Best}, {Baraffe},
  {Chabrier}, {Forveille}, {Metchev}, \& {Tremblin}}]{dupuy2020}
{Dupuy}, T.~J., {Liu}, M.~C., {Magnier}, E.~A., {et~al.} 2020, Research Notes
  of the American Astronomical Society, 4, 54, \dodoi{10.3847/2515-5172/ab8942}

\bibitem[{{Fabrycky} \& {Murray-Clay}(2010)}]{fabrycky2010}
{Fabrycky}, D.~C., \& {Murray-Clay}, R.~A. 2010, \apj, 710, 1408,
  \dodoi{10.1088/0004-637X/710/2/1408}

\bibitem[{{Faherty} {et~al.}(2013){Faherty}, {Rice}, {Cruz}, {Mamajek}, \&
  {N{\'u}{\~n}ez}}]{faherty2013}
{Faherty}, J.~K., {Rice}, E.~L., {Cruz}, K.~L., {Mamajek}, E.~E., \&
  {N{\'u}{\~n}ez}, A. 2013, \aj, 145, 2, \dodoi{10.1088/0004-6256/145/1/2}

\bibitem[{{Foreman-Mackey} {et~al.}(2013){Foreman-Mackey}, {Conley},
  {Meierjurgen Farr}, {Hogg}, {Lang}, {Marshall}, {Price-Whelan}, {Sanders}, \&
  {Zuntz}}]{foreman-mackey2013}
{Foreman-Mackey}, D., {Conley}, A., {Meierjurgen Farr}, W., {et~al.} 2013,
  {emcee: The MCMC Hammer}.
\newblock \doeprint{1303.002}

\bibitem[{{Galicher} {et~al.}(2016){Galicher}, {Marois}, {Macintosh},
  {Zuckerman}, {Barman}, {Konopacky}, {Song}, {Patience}, {Lafreni{\`e}re},
  {Doyon}, \& {Nielsen}}]{galicher2016}
{Galicher}, R., {Marois}, C., {Macintosh}, B., {et~al.} 2016, \aap, 594, A63,
  \dodoi{10.1051/0004-6361/201527828}

\bibitem[{{Gauza} {et~al.}(2015){Gauza}, {B{\'e}jar}, {P{\'e}rez-Garrido},
  {Zapatero Osorio}, {Lodieu}, {Rebolo}, {Pall{\'e}}, \& {Nowak}}]{gauza2015}
{Gauza}, B., {B{\'e}jar}, V. J.~S., {P{\'e}rez-Garrido}, A., {et~al.} 2015,
  \apj, 804, 96, \dodoi{10.1088/0004-637X/804/2/96}

\bibitem[{{Goodman} \& {Weare}(2010)}]{goodman2010}
{Goodman}, J., \& {Weare}, J. 2010, Communications in Applied Mathematics and
  Computational Science, 5, 65, \dodoi{10.2140/camcos.2010.5.65}

\bibitem[{{Hargreaves} {et~al.}(2020){Hargreaves}, {Gordon}, {Rey}, {Nikitin},
  {Tyuterev}, {Kochanov}, \& {Rothman}}]{hargreaves2020}
{Hargreaves}, R.~J., {Gordon}, I.~E., {Rey}, M., {et~al.} 2020, \apjs, 247, 55,
  \dodoi{10.3847/1538-4365/ab7a1a}

\bibitem[{{Hauschildt} {et~al.}(1999){Hauschildt}, {Allard}, \&
  {Baron}}]{hauschildt1999}
{Hauschildt}, P.~H., {Allard}, F., \& {Baron}, E. 1999, \apj, 512, 377,
  \dodoi{10.1086/306745}

\bibitem[{{Helled} \& {Schubert}(2009)}]{helled2009}
{Helled}, R., \& {Schubert}, G. 2009, \apj, 697, 1256,
  \dodoi{10.1088/0004-637X/697/2/1256}

\bibitem[{{Hinkley} {et~al.}(2017){Hinkley}, {Baraffe}, {Biller}, {Bonnefoy},
  {Bowler}, {Chen}, {Choquet}, {Currie}, {Danielski}, {Fortney}, {Grady},
  {Greenbaum}, {Hines}, {Janson}, {Kalas}, {Kennedy}, {Kraus}, {Lagrange},
  {Liu}, {Marley}, {Marois}, {Matthews}, {Mawet}, {Metchev}, {Meyer},
  {Millar-Blanchaer}, {Perrin}, {Pueyo}, {Quanz}, {Rameau}, {Rodigas},
  {Sallum}, {Sargent}, {Schlieder}, {Schneider}, {Skemer}, {Stapelfeldt},
  {Tremblin}, {Vigan}, \& {Ygouf}}]{hinkley2017}
{Hinkley}, S., {Baraffe}, I., {Biller}, B., {et~al.} 2017, {High Contrast
  Imaging of Exoplanets and Exoplanetary Systems with JWST}, JWST Proposal ID
  1386. Cycle 0 Early Release Science

\bibitem[{{Hsu} {et~al.}(2021{\natexlab{a}}){Hsu}, {Theissen}, {Burgasser}, \&
  {Birky}}]{hsu2021_smart}
{Hsu}, C.-C., {Theissen}, C., {Burgasser}, A., \& {Birky}, J.
  2021{\natexlab{a}}, {SMART: The Spectral Modeling Analysis and RV Tool},
  v1.0.0, Zenodo,  Zenodo, \dodoi{10.5281/zenodo.4765258}

\bibitem[{{Hsu} {et~al.}(2021{\natexlab{b}}){Hsu}, {Burgasser}, {Theissen},
  {Gelino}, {Birky}, {Diamant}, {Bardalez Gagliuffi}, {Aganze}, {Blake}, \&
  {Faherty}}]{hsu2021}
{Hsu}, C.-C., {Burgasser}, A.~J., {Theissen}, C.~A., {et~al.}
  2021{\natexlab{b}}, \apjs, 257, 45, \dodoi{10.3847/1538-4365/ac1c7d}

\bibitem[{{Husser} {et~al.}(2013){Husser}, {Wende-von Berg}, {Dreizler},
  {Homeier}, {Reiners}, {Barman}, \& {Hauschildt}}]{husser2013}
{Husser}, T.~O., {Wende-von Berg}, S., {Dreizler}, S., {et~al.} 2013, \aap,
  553, A6, \dodoi{10.1051/0004-6361/201219058}

\bibitem[{{Karkoschka} \& {Tomasko}(2010)}]{karkoschka2010}
{Karkoschka}, E., \& {Tomasko}, M.~G. 2010, \icarus, 205, 674,
  \dodoi{10.1016/j.icarus.2009.07.044}

\bibitem[{{Konopacky} {et~al.}(2013){Konopacky}, {Barman}, {Macintosh}, \&
  {Marois}}]{konopacky2013}
{Konopacky}, Q.~M., {Barman}, T.~S., {Macintosh}, B.~A., \& {Marois}, C. 2013,
  Science, 339, 1398, \dodoi{10.1126/science.1232003}

\bibitem[{{Krabbe} {et~al.}(2004){Krabbe}, {Gasaway}, {Song}, {Iserlohe},
  {Weiss}, {Larkin}, {Barczys}, \& {Lafreniere}}]{krabbe2004}
{Krabbe}, A., {Gasaway}, T., {Song}, I., {et~al.} 2004, in Society of
  Photo-Optical Instrumentation Engineers (SPIE) Conference Series, Vol. 5492,
  Ground-based Instrumentation for Astronomy, ed. A.~F.~M. {Moorwood} \&
  M.~{Iye}, 1403--1410, \dodoi{10.1117/12.552592}

\bibitem[{{Larkin} {et~al.}(2006){Larkin}, {Barczys}, {Krabbe}, {Adkins},
  {Aliado}, {Amico}, {Brims}, {Campbell}, {Canfield}, {Gasaway}, {Honey},
  {Iserlohe}, {Johnson}, {Kress}, {LaFreniere}, {Lyke}, {Magnone}, {Magnone},
  {McElwain}, {Moon}, {Quirrenbach}, {Skulason}, {Song}, {Spencer}, {Weiss}, \&
  {Wright}}]{larkin2006}
{Larkin}, J., {Barczys}, M., {Krabbe}, A., {et~al.} 2006, in Society of
  Photo-Optical Instrumentation Engineers (SPIE) Conference Series, Vol. 6269,
  Society of Photo-Optical Instrumentation Engineers (SPIE) Conference Series,
  ed. I.~S. {McLean} \& M.~{Iye}, 62691A, \dodoi{10.1117/12.672061}

\bibitem[{{Liu} {et~al.}(2021){Liu}, {Bitsch}, {Asplund}, {Liu}, {Murphy},
  {Yong}, {Ting}, \& {Feltzing}}]{liufan2021}
{Liu}, F., {Bitsch}, B., {Asplund}, M., {et~al.} 2021, \mnras, 508, 1227,
  \dodoi{10.1093/mnras/stab2471}

\bibitem[{{Liu} {et~al.}(2016){Liu}, {Dupuy}, \& {Allers}}]{liu2016}
{Liu}, M.~C., {Dupuy}, T.~J., \& {Allers}, K.~N. 2016, \apj, 833, 96,
  \dodoi{10.3847/1538-4357/833/1/96}

\bibitem[{{Lockhart} {et~al.}(2019){Lockhart}, {Do}, {Larkin}, {Boehle},
  {Campbell}, {Chappell}, {Chu}, {Ciurlo}, {Cosens}, {Fitzgerald}, {Ghez},
  {Lu}, {Lyke}, {Mieda}, {Rudy}, {Vayner}, {Walth}, \& {Wright}}]{lockhart2019}
{Lockhart}, K.~E., {Do}, T., {Larkin}, J.~E., {et~al.} 2019, \aj, 157, 75,
  \dodoi{10.3847/1538-3881/aaf64e}

\bibitem[{{Madhusudhan}(2019)}]{madhu2019}
{Madhusudhan}, N. 2019, \araa, 57, 617,
  \dodoi{10.1146/annurev-astro-081817-051846}

\bibitem[{{McMahon} {et~al.}(2013){McMahon}, {Banerji}, {Gonzalez}, {Koposov},
  {Bejar}, {Lodieu}, {Rebolo}, \& {VHS Collaboration}}]{mcmahon2013}
{McMahon}, R.~G., {Banerji}, M., {Gonzalez}, E., {et~al.} 2013, The Messenger,
  154, 35

\bibitem[{{Miles} {et~al.}(2018){Miles}, {Skemer}, {Barman}, {Allers}, \&
  {Stone}}]{miles2018}
{Miles}, B.~E., {Skemer}, A.~J., {Barman}, T.~S., {Allers}, K.~N., \& {Stone},
  J.~M. 2018, \apj, 869, 18, \dodoi{10.3847/1538-4357/aae6cd}

\bibitem[{{Molli{\`e}re} {et~al.}(2020){Molli{\`e}re}, {Stolker}, {Lacour},
  {Otten}, {Shangguan}, {Charnay}, {Molyarova}, {Nowak}, {Henning}, {Marleau},
  {Semenov}, {van Dishoeck}, {Eisenhauer}, {Garcia}, {Garcia Lopez}, {Girard},
  {Greenbaum}, {Hinkley}, {Kervella}, {Kreidberg}, {Maire}, {Nasedkin},
  {Pueyo}, {Snellen}, {Vigan}, {Wang}, {de Zeeuw}, \& {Zurlo}}]{molliere2020}
{Molli{\`e}re}, P., {Stolker}, T., {Lacour}, S., {et~al.} 2020, \aap, 640,
  A131, \dodoi{10.1051/0004-6361/202038325}

\bibitem[{{Nielsen} {et~al.}(2019){Nielsen}, {De Rosa}, {Macintosh}, {Wang},
  {Ruffio}, {Chiang}, {Marley}, {Saumon}, {Savransky}, {Ammons}, {Bailey},
  {Barman}, {Blain}, {Bulger}, {Burrows}, {Chilcote}, {Cotten}, {Czekala},
  {Doyon}, {Duch{\^e}ne}, {Esposito}, {Fabrycky}, {Fitzgerald}, {Follette},
  {Fortney}, {Gerard}, {Goodsell}, {Graham}, {Greenbaum}, {Hibon}, {Hinkley},
  {Hirsch}, {Hom}, {Hung}, {Dawson}, {Ingraham}, {Kalas}, {Konopacky},
  {Larkin}, {Lee}, {Lin}, {Maire}, {Marchis}, {Marois}, {Metchev},
  {Millar-Blanchaer}, {Morzinski}, {Oppenheimer}, {Palmer}, {Patience},
  {Perrin}, {Poyneer}, {Pueyo}, {Rafikov}, {Rajan}, {Rameau}, {Rantakyr{\"o}},
  {Ren}, {Schneider}, {Sivaramakrishnan}, {Song}, {Soummer}, {Tallis},
  {Thomas}, {Ward-Duong}, \& {Wolff}}]{nielsen2019}
{Nielsen}, E.~L., {De Rosa}, R.~J., {Macintosh}, B., {et~al.} 2019, \aj, 158,
  13, \dodoi{10.3847/1538-3881/ab16e9}

\bibitem[{{Nikolov} {et~al.}(2018){Nikolov}, {Sing}, {Goyal}, {Henry},
  {Wakeford}, {Evans}, {L{\'o}pez-Morales}, {Garc{\'\i}a Mu{\~n}oz},
  {Ben-Jaffel}, {Sanz-Forcada}, {Ballester}, {Kataria}, {Barstow}, {Bourrier},
  {Buchhave}, {Cohen}, {Deming}, {Ehrenreich}, {Knutson}, {Lavvas}, {Lecavelier
  des Etangs}, {Lewis}, {Mandell}, \& {Williamson}}]{nikolov2018}
{Nikolov}, N., {Sing}, D.~K., {Goyal}, J., {et~al.} 2018, \mnras, 474, 1705,
  \dodoi{10.1093/mnras/stx2865}

\bibitem[{{{\"O}berg} {et~al.}(2011){{\"O}berg}, {Murray-Clay}, \&
  {Bergin}}]{oberg2011}
{{\"O}berg}, K.~I., {Murray-Clay}, R., \& {Bergin}, E.~A. 2011, \apjl, 743,
  L16, \dodoi{10.1088/2041-8205/743/1/L16}

\bibitem[{{Oreshenko} {et~al.}(2020){Oreshenko}, {Kitzmann},
  {M{\'a}rquez-Neila}, {Malik}, {Bowler}, {Burgasser}, {Sznitman}, {Fisher}, \&
  {Heng}}]{oreshenko2020}
{Oreshenko}, M., {Kitzmann}, D., {M{\'a}rquez-Neila}, P., {et~al.} 2020, \aj,
  159, 6, \dodoi{10.3847/1538-3881/ab5955}

\bibitem[{{Petrus} {et~al.}(2021){Petrus}, {Bonnefoy}, {Chauvin}, {Charnay},
  {Marleau}, {Gratton}, {Lagrange}, {Rameau}, {Mordasini}, {Nowak}, {Delorme},
  {Boccaletti}, {Carlotti}, {Houll{\'e}}, {Vigan}, {Allard}, {Desidera},
  {D'Orazi}, {Hoeijmakers}, {Wyttenbach}, \& {Lavie}}]{petrus2021}
{Petrus}, S., {Bonnefoy}, M., {Chauvin}, G., {et~al.} 2021, \aap, 648, A59,
  \dodoi{10.1051/0004-6361/202038914}

\bibitem[{{Rameau} {et~al.}(2013){Rameau}, {Chauvin}, {Lagrange}, {Meshkat},
  {Boccaletti}, {Quanz}, {Currie}, {Mawet}, {Girard}, {Bonnefoy}, \&
  {Kenworthy}}]{rameau2013}
{Rameau}, J., {Chauvin}, G., {Lagrange}, A.~M., {et~al.} 2013, \apjl, 779, L26,
  \dodoi{10.1088/2041-8205/779/2/L26}

\bibitem[{{Rich} {et~al.}(2016){Rich}, {Currie}, {Wisniewski}, {Hashimoto},
  {Brandt}, {Carson}, {Kuzuhara}, \& {Uyama}}]{rich2016}
{Rich}, E.~A., {Currie}, T., {Wisniewski}, J.~P., {et~al.} 2016, \apj, 830,
  114, \dodoi{10.3847/0004-637X/830/2/114}

\bibitem[{{Ruffio} {et~al.}(2021){Ruffio}, {Konopacky}, {Barman}, {Macintosh},
  {Wilcomb}, {De Rosa}, {Wang}, {Czekala}, \& {Marois}}]{ruffio2021}
{Ruffio}, J.-B., {Konopacky}, Q.~M., {Barman}, T., {et~al.} 2021, arXiv
  e-prints, arXiv:2109.07614.
\newblock \doarXiv{2109.07614}

\bibitem[{{Skrutskie} {et~al.}(2006){Skrutskie}, {Cutri}, {Stiening},
  {Weinberg}, {Schneider}, {Carpenter}, {Beichman}, {Capps}, {Chester},
  {Elias}, {Huchra}, {Liebert}, {Lonsdale}, {Monet}, {Price}, {Seitzer},
  {Jarrett}, {Kirkpatrick}, {Gizis}, {Howard}, {Evans}, {Fowler}, {Fullmer},
  {Hurt}, {Light}, {Kopan}, {Marsh}, {McCallon}, {Tam}, {Van Dyk}, \&
  {Wheelock}}]{skrutskie2006}
{Skrutskie}, M.~F., {Cutri}, R.~M., {Stiening}, R., {et~al.} 2006, \aj, 131,
  1163, \dodoi{10.1086/498708}

\bibitem[{{Stone} {et~al.}(2016){Stone}, {Skemer}, {Kratter}, {Dupuy}, {Close},
  {Eisner}, {Fortney}, {Hinz}, {Males}, {Morley}, {Morzinski}, \&
  {Ward-Duong}}]{stone2016}
{Stone}, J.~M., {Skemer}, A.~J., {Kratter}, K.~M., {et~al.} 2016, \apjl, 818,
  L12, \dodoi{10.3847/2041-8205/818/1/L12}

\bibitem[{{Theissen} {et~al.}(2022){Theissen}, {Konopacky}, {Lu}, {Kim},
  {Zhang}, {Hsu}, {Chu}, \& {Wei}}]{theissen2022}
{Theissen}, C.~A., {Konopacky}, Q.~M., {Lu}, J.~R., {et~al.} 2022, \apj, 926,
  141, \dodoi{10.3847/1538-4357/ac3252}

\bibitem[{{Vigan} {et~al.}(2021){Vigan}, {Fontanive}, {Meyer}, {Biller},
  {Bonavita}, {Feldt}, {Desidera}, {Marleau}, {Emsenhuber}, {Galicher}, {Rice},
  {Forgan}, {Mordasini}, {Gratton}, {Le Coroller}, {Maire}, {Cantalloube},
  {Chauvin}, {Cheetham}, {Hagelberg}, {Lagrange}, {Langlois}, {Bonnefoy},
  {Beuzit}, {Boccaletti}, {D'Orazi}, {Delorme}, {Dominik}, {Henning}, {Janson},
  {Lagadec}, {Lazzoni}, {Ligi}, {Menard}, {Mesa}, {Messina}, {Moutou},
  {M{\"u}ller}, {Perrot}, {Samland}, {Schmid}, {Schmidt}, {Sissa}, {Turatto},
  {Udry}, {Zurlo}, {Abe}, {Antichi}, {Asensio-Torres}, {Baruffolo}, {Baudoz},
  {Baudrand}, {Bazzon}, {Blanchard}, {Bohn}, {Brown Sevilla}, {Carbillet},
  {Carle}, {Cascone}, {Charton}, {Claudi}, {Costille}, {De Caprio},
  {Delboulb{\'e}}, {Dohlen}, {Engler}, {Fantinel}, {Feautrier}, {Fusco},
  {Gigan}, {Girard}, {Giro}, {Gisler}, {Gluck}, {Gry}, {Hubin}, {Hugot},
  {Jaquet}, {Kasper}, {Le Mignant}, {Llored}, {Madec}, {Magnard}, {Martinez},
  {Maurel}, {M{\"o}ller-Nilsson}, {Mouillet}, {Moulin}, {Orign{\'e}}, {Pavlov},
  {Perret}, {Petit}, {Pragt}, {Puget}, {Rabou}, {Ramos}, {Rickman}, {Rigal},
  {Rochat}, {Roelfsema}, {Rousset}, {Roux}, {Salasnich}, {Sauvage}, {Sevin},
  {Soenke}, {Stadler}, {Suarez}, {Wahhaj}, {Weber}, \& {Wildi}}]{vigan2021}
{Vigan}, A., {Fontanive}, C., {Meyer}, M., {et~al.} 2021, \aap, 651, A72,
  \dodoi{10.1051/0004-6361/202038107}

\bibitem[{{Vos} {et~al.}(2019){Vos}, {Biller}, {Bonavita}, {Eriksson}, {Liu},
  {Best}, {Metchev}, {Radigan}, {Allers}, {Janson}, {Buenzli}, {Dupuy},
  {Bonnefoy}, {Manjavacas}, {Brandner}, {Crossfield}, {Deacon}, {Henning},
  {Homeier}, {Kopytova}, \& {Schlieder}}]{vos2019}
{Vos}, J.~M., {Biller}, B.~A., {Bonavita}, M., {et~al.} 2019, \mnras, 483, 480,
  \dodoi{10.1093/mnras/sty3123}

\bibitem[{{Wang} {et~al.}(2022){Wang}, {Kolecki}, {Ruffio}, {Wang}, {Mawet},
  {Baker}, {Bartos}, {Blake}, {Bond}, {Calvin}, {Cetre}, {Delorme}, {Doppmann},
  {Echeverri}, {Finnerty}, {Fitzgerald}, {Jovanovic}, {Liu}, {Lopez}, {Morris},
  {Pai Asnodkar}, {Pezzato}, {Ragland}, {Roy}, {Ruane}, {Sappey}, {Schofield},
  {Skemer}, {Venenciano}, {Kent Wallace}, {Wallack}, {Wizinowich}, \&
  {Xuan}}]{wang2022}
{Wang}, J., {Kolecki}, J.~R., {Ruffio}, J.-B., {et~al.} 2022, \aj, 163, 189,
  \dodoi{10.3847/1538-3881/ac56e2}

\bibitem[{{Wilcomb} {et~al.}(2020){Wilcomb}, {Konopacky}, {Barman}, {Theissen},
  {Ruffio}, {Brock}, {Macintosh}, \& {Marois}}]{wilcomb2020}
{Wilcomb}, K.~K., {Konopacky}, Q.~M., {Barman}, T.~S., {et~al.} 2020, \aj, 160,
  207, \dodoi{10.3847/1538-3881/abb9b1}

\bibitem[{{Zhang} {et~al.}(2021){Zhang}, {Snellen}, {Bohn}, {Molli{\`e}re},
  {Ginski}, {Hoeijmakers}, {Kenworthy}, {Mamajek}, {Meshkat}, {Reggiani}, \&
  {Snik}}]{zhang2021}
{Zhang}, Y., {Snellen}, I. A.~G., {Bohn}, A.~J., {et~al.} 2021, \nat, 595, 370,
  \dodoi{10.1038/s41586-021-03616-x}

\end{thebibliography}

\end{document}